\documentclass[journal]{IEEEtran}
\ifCLASSINFOpdf
\else
\fi

\usepackage[utf8]{inputenc}
\usepackage{amsmath,amssymb,amsfonts}
\usepackage{cite}

\usepackage{algorithmic}
\usepackage{stfloats}
\usepackage{graphicx}
\usepackage{textcomp}
\usepackage{xcolor}
\usepackage{color}
\usepackage{pifont}
\usepackage{amsthm}

\usepackage{hyperref}
\usepackage{mathrsfs}
\usepackage{booktabs}
\usepackage{hyperref}
\usepackage{cases}
\usepackage{comment}
\usepackage{empheq} 
\usepackage{caption}
\usepackage{subcaption}
\allowdisplaybreaks[4]

\usepackage{url}  
\usepackage{graphicx}  
\usepackage[ruled,vlined,linesnumbered]{algorithm2e}

\begin{document}
%
\title{Secure Enhancement for RIS-Aided UAV with ISAC: Robust Design and Resource Allocation}
%
%
%

\author{Yue Xiu,~Wanting Lyu,~Phee~Lep Yeoh,~\IEEEmembership{Senior Member,~IEEE}\\~Yonghui Li,~\IEEEmembership{Fellow,~IEEE}, Yi Ai,
~Ning Wei,~\IEEEmembership{Member,~IEEE}\\
\thanks{
Y.~Xiu and W.~Lyu, N.~Wei are with National Key Laboratory of Science and Technology on Communications, University of Electronic Science and Technology of China, Chengdu 611731, China (E-mail: xiuyue12345678@163.com, 
lyuwanting@yeah.net, wn@uestc.edu.cn).
Phee Lep Yeoh is with School of Science, Technology and Engineering, University of the Sunshine Coast, QLD, Australia(e-mail: pyeoh@usc.edu.au).
Yonghui Li is with the School of Electrical and Information Engineering, University of Sydney, Sydney, NSW 2006, Australia(e-mail: yonghui.li@sydney.edu.au).
Yi Ai is with School of Science, Technology and Engineering, Southwest Jiaotong University, Cheng Du, China (e-mail: aiyi@my.swjtu.edu.cn).
}

\thanks{The corresponding author is Ning Wei.}}

\maketitle

\begin{abstract}
This paper analyses the security performance of a reconfigurable intelligent surface (RIS)-aided unmanned aerial vehicle (UAV) communication system with integrated sensing and communications (ISAC). We consider a multiple-antenna UAV transmitting ISAC waveforms to simultaneously detect an untrusted target in the surrounding environment and communicate with a ground Internet-of-Things (IoT) device in the presence of an eavesdropper (Eve). Given that the Eve can conceal their channel state information (CSI) in practical scenarios, we assume that the CSI of the eavesdropper channel is imperfect. For this RIS-aided ISAC-UAV system, we aim to maximize the average communication secrecy rate by jointly optimizing UAV trajectory, RIS passive beamforming, transmit beamforming, and receive beamforming. However, this joint optimization problem is non-convex due to multi-variable coupling. As such, we solve the optimization using an efficient and tractable algorithm using a block coordinate descent (BCD) method. Specifically, we develop a successive convex approximation (SCA) algorithm based on semidefinite relaxation (SDR) to optimise the joint optimization as four separate non-convex subproblems. Numerical results show that our proposed algorithm can successfully ensure the accuracy of sensing targets and significantly improve the communication secrecy rate of the IoT communication devices. 
\end{abstract}


%
\IEEEpeerreviewmaketitle

\IEEEPARstart{I}{n} recent years, with the development of autonomous driving and artificial intelligence, demands for data traffic in wireless communications have increased exponentially, which poses a significant challenge for the development of wireless communication systems. Future generation 6G wireless systems will need to support secure low-latency communications to various advanced applications including the emerging Internet-of-Things (IoT) networks. An attractive solution to provide flexible coverage to massive numbers of ground IoT devices is the deployment of unmanned aerial vehicle (UAV) base stations (BS)\cite{8873672,8456560}.  Compared with a conventional fixed BS, UAVs are not restricted by terrain and can be deployed nearer to the ground IoT devices to improve communication quality using line-of-sight (LoS) transmission. Despite these advantages, communication security emerges as a notable challenge of UAV communications since other untrusted devices could be directly eavesdropping on the UAV's LoS transmissions \cite{9954169,8432499,9210742,9410619}. In \cite{9954169}, the authors proposed a novel online edge learning offloading (OELO) scheme for UAV-assisted MEC secure communication, which can improve secure communication performance. In\cite{8432499}, the authors analysed the secrecy performance of a UAV communication and jamming system with joint trajectory optimization and user scheduling. In\cite{9410619}, the authors studied the security performance of UAV-to-vehicle communication systems. They derived approximate and closed-form expressions for the asymptotic security outage probability (SOP) of downlinks experiencing Rician fading channels.

A major constraint for UAV communications is that the LoS transmission signals could be blocked. Recently, the use of reconfigurable intelligent surface (RIS) has been proposed to enhance UAV communications. Specifically, in\cite{9124704}, the author analysed the achievable rate of a RIS-assisted UAV relay system. In\cite{8959174}, the authors used UAVs as mobile BSs to provide services to ground users with the help of RIS. In\cite{9013626}, the author proposed a deep reinforcement learning algorithm to solve the system's communication rate where RIS is placed on UAVs.

In advanced 6G IoT applications, the BS may also need to provide target sensing functions to obtain IoT location information as well as communication functions which has resulted in the emergence of a new wireless communication concept, namely integrated sensing and communications (ISAC)\cite{9606831}. The integration gains brought by the cooperation of sensing and communications have attracted widespread attention from academia and industry in\cite{9737357,5440129,10158711}.
In \cite{9737357,5440129}, three different approaches to ISAC are considered, namely, communication-centered design, radar-centered design, and jointly designed ISAC technology. Among them, communication-centered design aims to use communication waveforms to meet sensing needs whilst radar-centered design aims to create radar waveforms to meet communication transmission needs. Joint design aims to create new waveforms to meet both communication and sensing needs. In\cite{10158711}, the authors use extended existing ISAC beamforming designs to the general case by considering Full-Duplex (FD) capabilities of radar and communications and designed an optimal receive beamformer for transmit beamforming and user transmit power determination. Although there are many advantages of ISAC systems in improving communication transmission rates, there are still problems with ISAC technology regarding communication security. In ISAC technology, the BS beam must emit beams to sense the target. However, if the sensing target is also an eavesdropper (Eve), it can eavesdrop on potentially sensitive user information. Therefore, the secure communication of ISAC technology is a critical research topic\cite{9927490,10153696,9933849}. In\cite{9927490}, the authors utilized non-orthogonal multiple access (NOMA) to support more users of the ISAC network and ensure security through carefully designed precoding.
In\cite{10153696}, the authors studied the downlink security integrated ISAC system and considered the bounded error model of Eve's CSI, where the worst-case confidentiality rate constraint is adopted to ensure secure communication performance.
In\cite{9933849}, the authors studied the design of robust resource allocation for secure communication in ISAC systems.  Moreover, several studies \cite{10168298,li2024uav,10529955,son2024secrecy,10054167} have explored Physical Layer Security (PLS) techniques in ISAC-UAV systems, where the flexibility of UAVs offers increased degrees of freedom (DoFs) for enhancing security and sensing performance. For instance, in \cite{10168298,li2024uav}, the authors proposed an semidefinite programming (SDP) relaxation to interfere with a ground eavesdropper (Eve) while improving target sensing accuracy and overall secrecy rates. Additionally, in \cite{10529955,son2024secrecy}, the authors employed the extended Kalman filter technique to predict user motion states, enabling the joint design of radar signals and receiver beamforming to enhance the secrecy rate. In \cite{10054167}, the authors introduced a novel adaptive ISAC for UAV systems, allowing UAVs to sense on-demand during communication processes. The sensing duration is flexibly configured based on application requirements, thus avoiding over-sensing and radio resource waste while improving resource utilization and system secrecy rate.

In this paper, we propose a new framework to enhance the security of RIS-aided UAV BSs with ISAC operating in the presence of Eve and an untrusted target (UT). Different from the existing ISAC-UAV research which only considers Eve or UT\cite{10168298,li2024uav,10529955}, this paper considers a more general scenario, that is, the coexistence of Eve and UT. 
We define, Eve as an active eavesdropper that can detect and intercept the communication information between the UAV and IoT device whereas the UT is a passive eavesdropper, which can only intercept information when the UAV is sensing the target\cite{10334744,10153696}. We note that in most previous works on secure ISAC-UAV communication systems\cite{son2024secrecy,10054167}, the CSI of Eve and UT is assumed to be perfectly known, and the sensing target is considered to have no communication security problems. This assumption may be challenging in practice where the Eve and UT can try to avoid detection by the UAV, which results in uncertainty in their CSI. 
Inspired by this, we consider a RIS-assisted ISAC-UAV secure communication system, in which UAVs provide services to ground IoT devices and UT, whilst the UT and Eve attempt to eavesdrop on the communication of the IoT device. We assume that Eve's CSI and UT's CSI are imperfect and use a deterministic model\cite{10158711} to describe the CSI uncertainty. Based on this, the contributions of this paper are summarized as follows:
\begin{itemize}
\item
We analyse the security performance of RIS-assisted ISAC-UAV systems in the presence of both Eve and an untrusted target (UT). For this system, we formulate a robust optimization of the ISAC-UAV's trajectory, the RIS passive beamforming, and transmit/receive beamforming to maximize the average worst-case secrecy rate. Due to the coupling of these optimization variables and the assumption of imperfect CSI for Eve and UT, this optimization problem is non-convex. To solve this problem, we propose an efficient algorithm based on the block coordinate descent (BCD) method.
\item
Based on the BCD method, the original problem is equivalently transformed as four sub-problems: \textbf{(1)} transmit beamforming optimization, \textbf{(2)} RIS beamforming optimization, \textbf{(3)} UAV trajectory optimization, and \textbf{(4)} UAV receive beamforming optimization. Specifically, for subproblem 1, we calculate the optimal transmit beamforming based on the objective function's unique structure using the triangle inequality. For sub-problem 2, we apply the successive convex approximation (SCA) and semidefinite relaxation (SDR) methods to deal with CSI uncertainty and RIS beamforming design issues. For sub-problem 3, we use the SCA method to solve with the UAV trajectory optimization problem. For sub-problem 4, we use the Majorize-Minimization(MM) algorithm to solve the receive beamforming optimization problem.
\item
Simulation results show the proposed algorithm can achieve significant improvements in the secrecy rate with increasing UAV transmit power. Our results confirm that the approximation can be sufficiently accurate even using only tens of RIS reflectors. We also highlight the tradeoff between satisfying the secrecy rate and the sensing rate in the optimized ISAC-UAV trajectory.
\end{itemize}

\textbf{Organization:} This paper contains the following sections. In section \ref{II}, the system model and problem formulation are introduced. In section \ref{III}, the proposed algorithm is presented. Specifically, this section includes transmitting beamforming design, RIS reflection beamforming design, and parameter optimization. Section \ref{IV} analyzes the convergence of the proposed algorithm. Section \ref{V} proves the performance of the proposed algorithm by using numerical results. Section \ref{VI} summarizes the paper.

\section{System Model and Problem Formulation}\label{II}
\begin{figure}[!h]
  \centering
  \includegraphics[scale=0.4]{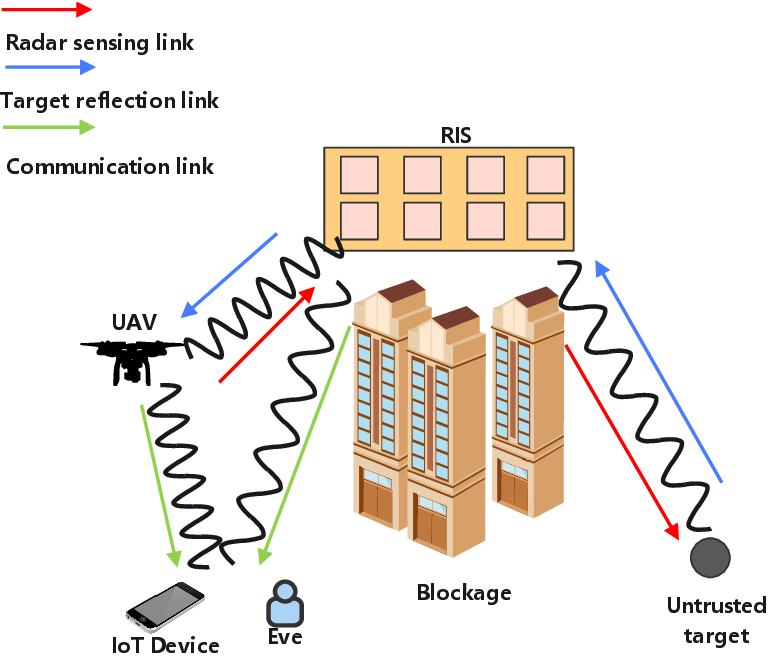}
  \caption{Illustation of RIS-aided ISAC-UAV System sending communications to the IoT device in the presence of an Eve and sensing the location of the untrusted target (UT). The UT is also considered as another eavesdropper attempting to access the communications sent to the IoT device.}
\label{FIGURE0}
\end{figure}
We consider a secure ISAC-UAV system, as shown in Fig.\ref{FIGURE0}. In this system, the ISAC-UAV is equipped with a uniform linear array (ULA) and has $N_{t}$ transmitting antennas and sensing receiving antennas. The locations of the IoT device, UT, Eve, and RIS are expressed according to a 3D coordinate system as $\mathbf{p}_{d}=[x_{d}, y_{d},0]^{T}$, $\mathbf{p}_{t}[s]=[x_{t}[s], y_{t}[s],0]^{T}$and $\mathbf{p}_{e}[s]=[x_{e}[s], y_{e}[s], 0]^{T}$ and $p_{r} = [x_{r}, y_{r}, z_{r}]$, respectively. To perform the UAV trajectory optimization, consider the UAV needs to complete the ISAC functions within a given time duration of $T$ which is divided into $S$ slots, where $T=S\Delta_{s}$ and $\Delta_{s}$ represent the slot length, $s\in\{1,\ldots,S\}$ is the $s$-th time slot. The position of the UAV is expressed as $\mathbf{p}_{u}[s]=[x_{u}[s], y_{u}[s], z_{u}]^{T}$, where the UAV is flying horizontally, so the height of the UAV remains the same. We assume that the direct link between the UAV and the UT is blocked. Thus, the signal transmitted from the UAV directly reaches the user and the perceived target through RIS to the user and detects the target simultaneously. When the signal is emitted, the signal reflected by the RIS acts on the target and is reflected to the RIS by the target. Finally, the signal is reflected to the UAV. According to \cite{9737357}, we assume that the UAV can successfully eliminate the self-interference of the transmitted signal and the received echo.

\subsection{Communication Signal Model}
For the communication link, the ISAC signal $x[s]$ sent by the UAV reaches the IoT device and Eve through the RIS reflection link and direct link in each $s$-th time slot. The signals received at the IoT device, Eve and UT, are respectively expressed as
\begin{small}
\begin{align}
&y_{ud}[s]=\left(L_{ud}[s]\mathbf{h}_{ud}^{H}[s]+L_{urd}[s]\mathbf{h}_{rd}^{H}\boldsymbol{\Theta}[s]\mathbf{H}_{ur}[s]\right)\mathbf{w}[s]x[s]+n_{d}[s],\label{pro1}
\end{align}   
\end{small}%
and
\begin{small}
\begin{align}
&y_{ue}[s]=\left(L_{ue}[s]\mathbf{h}_{ue}^{H}[s]+L_{ure}[s]\mathbf{h}_{re}^{H}\boldsymbol{\Theta}[s]\mathbf{H}_{ur}[s]\right)\mathbf{w}[s]x[s]+n_{e}[s],\label{pro2}
\end{align}
\end{small}%
and
\begin{small}
\begin{align}
&y_{ut}[s]=\left(L_{urt}[s]\mathbf{h}_{rt}^{H}\boldsymbol{\Theta}[s]\mathbf{H}_{ur}[s]\right)\mathbf{w}[s]x[s]+n_{t}[s],\label{pro3}
\end{align}
\end{small}%
where $L_{ud}[s]=\sqrt{\rho(z_{u}^{2}+\|\bar{\mathbf{p}}_{u}[s]-\mathbf{p}_{d}\|)^{-\frac{\kappa}{2}}}$, $L_{urd}[s]=\sqrt{\rho(d_{ur}[s]d_{rd})^{-\alpha}}$, $L_{ue}[s]=\sqrt{\rho(z_{u}^{2}+}$\\ $\overline{\|\bar{\mathbf{p}}_{u}[s]-\mathbf{p}_{e}[s]\|)^{-\frac{\kappa}{2}}}$, $L_{urt}[s]=\sqrt{\rho(d_{ur}[s]d_{rt})^{-\alpha}}$ and $L_{ure}[s]$ $=\sqrt{\rho(d_{ur}[s]d_{re})^{-\alpha}}$ denote the distance-dependent path loss model of the links from the UAV-to-IoT device, UAV-to-RIS-to-IoT device, UAV-to-Eve, UAV-to-RIS-UT, UAV-to-RIS-to-Eve, respectively\cite{9117136}, where 
$d_{ur}[s]=\sqrt{(z_{u}-z_{r})^{2}+\|\bar{\mathbf{p}}_{u}[s]-\mathbf{p}_{r}\|^{2}}$, $
d_{rd}=$
$\sqrt{z_{r}^{2}+\|\mathbf{p}_{r}-\mathbf{p}_{d}\|^{2}}$, $
d_{re}=\sqrt{z_{r}^{2}+\|\mathbf{p}_{r}-\mathbf{w}_{e}\|^{2}}$, $
d_{rt}=\sqrt{z_{r}^{2}+\|\mathbf{p}_{r}-\mathbf{p}_{t}\|^{2}}$. $\rho$ is the path loss at the reference distance $1$m. $\alpha$ and $\kappa$ are the path loss exponents for the UAV-to-IoT device link, UAV-to-RIS-to-IoT device link, the UAV-to-Eve link, the UAV-to-RIS-Eve link and the UAV-to-RIS-UT link, respectively.
According to \cite{4667717}, we assume that the channel models of all communication links are Rician channel models. Hence, the small-scale fading channel models $\mathbf{H}_{ur}[s]=\sqrt{\frac{\beta_{ur}}{1+\beta_{ur}}}\mathbf{H}_{ur}^{LoS}[s]+\sqrt{\frac{1}{1+\beta_{ur}}}\mathbf{H}_{ur}^{NLoS}[s]$, $
\mathbf{h}_{ud}[s]=$ $\sqrt{\frac{\beta_{ud}}{1+\beta_{ud}}}\mathbf{h}_{ud}^{LoS}[s]+\sqrt{\frac{1}{1+\beta_{ur}}}\mathbf{h}_{ud}^{NLoS}[s]$, $
\mathbf{h}_{rd}[s]=\sqrt{\frac{\beta_{rd}}{1+\beta_{rd}}}\mathbf{h}_{rd}^{LoS}[s]$ $+\sqrt{\frac{1}{1+\beta_{rd}}}\mathbf{h}_{rd}^{NLoS}[s]$, $\mathbf{h}_{ue}[s]=\sqrt{\frac{\beta_{ue}}{1+\beta_{ue}}}\mathbf{h}_{ue}^{LoS}[s]+\sqrt{\frac{1}{1+\beta_{ue}}}$ $\mathbf{h}_{ue}^{NLoS}[s]$,
$\mathbf{h}_{re}[s]=\sqrt{\frac{\beta_{re}}{1+\beta_{re}}}\mathbf{h}_{re}^{LoS}[s]+\sqrt{\frac{1}{1+\beta_{re}}}\mathbf{h}_{re}^{NLoS}[s]$, $\mathbf{h}_{rt}[s]=\sqrt{\frac{\beta_{rt}}{1+\beta_{rt}}}\mathbf{h}_{rt}^{LoS}[s]+\sqrt{\frac{1}{1+\beta_{rt}}}\mathbf{h}_{rt}^{NLoS}[s]$, where $\beta_{ur}$, $\beta_{ud}$, $\beta_{rd}$, $\beta_{ue}$, $\beta_{re}$, $\beta_{rt}$, $\beta_{ut}$ are the Rician factor of the UAV-to-RIS, UAV-to-IoT device, RIS-to-IoT device, UAV-to-Eve, RIS-to-Eve, and RIS-to-UT, respectively. $\mathbf{H}_{ur}^{LoS}[s]$, $\mathbf{h}_{ud}^{LoS}[s]$, $\mathbf{h}_{rd}^{LoS}[s]$, $\mathbf{h}_{ue}^{LoS}[s]$, $\mathbf{h}_{re}^{LoS}[s]$, $\mathbf{h}_{rt}^{LoS}[s]$ denote the channel LoS component. 
$\mathbf{H}_{ur}^{NLoS}[s]$, $\mathbf{h}_{ud}^{NLoS}[s]$, $\mathbf{h}_{rd}^{NLoS}[s]$, $\mathbf{h}_{ue}^{NLoS}[s]$, $\mathbf{h}_{re}^{NLoS}[s]$, $\mathbf{h}_{rt}^{NLoS}[s]$ denote the channel NLoS component.
$\boldsymbol{\Theta}[s]\in\mathbb{C}^{N_{R}\times N_{R}}=\mathrm{diag}\left(e^{j\theta_{1}[s]},e^{j\theta_{2}[s]},\ldots,\right.$ $\left.e^{j\theta_{N_{R}}[s]}\right)$ denotes the RIS reflection matrix, where $\theta_{i}[s],i\in[1,\ldots,N_{R}]$ is the $i$-th RIS reflector and $N_{R}$ is the number of RIS reflectors.

\subsection{Target Sensing Model}
In addition to transmitting the information from the UAV to the IoT device, $x[s]$ is also used for target detection. As shown in Fig.~\ref{FIGURE0}, signal $x[s]$ goes through the UAV-to-RIS-to-target-to-RIS-to-UAV link. Finally, the reflected echo signal is collected at the UAV. The mathematical expression of the echo signal is as follows
\begin{small}
\begin{align}
&r_{ut}[s]=\mathbf{u}^{H}L_{urt}^{2}[s]\mathbf{H}_{ur}[s]^{H}\boldsymbol{\Theta}[s]\mathbf{A}_{rt}\boldsymbol{\Theta}[s]\mathbf{H}_{ur}[s]\mathbf{w}[s]x[s]+n_{ut}[s],\label{pro4}
\end{align}    
\end{small}%
where $\mathbf{u}\in\mathbb{C}^{N_{t}\times1}$ is the receive beamformer at the UAV and $\|\mathbf{u}\|_{2}=1$. $n_{ut}[s]\sim\mathcal{CN}(0,\sigma^{2}_{ut})$ is the AWGN at the UAV. $\mathbf{A}_{rt}$ denotes the target response matrix at the RIS\cite{10306322}, and $\mathbf{A}_{rt}=\alpha_{s}\mathbf{a}\mathbf{a}^{H}$, where $\alpha_{s}$ is the RIS-to-UT-to-RIS path gain; $\mathbf{a}$ is the steering vector.

\subsection{Performance Metrics of Communication, Security and Sensing}
To analyse the performance of the communications link, we derive the SNR of the IoT device based on (\ref{pro1}), which results in
\begin{small}
\begin{align}
&\mathrm{SNR}_{ud}[s]=\frac{|L_{ud}[s]\mathbf{h}_{ud}^{H}[s]\mathbf{w}[s]+L_{udr}[s]\mathbf{h}_{rd}^{H}\boldsymbol{\Theta}[s]\mathbf{H}_{ur}[s]\mathbf{w}[s]|^{2}}{\sigma^{2}}.\label{pro6}
\end{align}    
\end{small}%
Based on (\ref{pro2}), the SNR of the Eve is derived as
\begin{small}
\begin{align}
&\mathrm{SNR}_{ue}[s]=\frac{|L_{ue}[s]\mathbf{h}_{ue}^{H}[s]\mathbf{w}[s]+L_{ure}[s]\mathbf{h}_{re}^{H}\boldsymbol{\Theta}[s]\mathbf{H}_{ur}[s]\mathbf{w}[s]|^{2}}{\sigma^{2}}.\label{pro7}
\end{align}
\end{small}%
Next, we apply (\ref{pro6}) and (\ref{pro7}) to obtain the communication secrecy rate of the IoT device with respect to Eve is given by
\begin{small}
\begin{align}
&R_{sec}[s]=\log_{2}(1+\mathrm{SNR}_{ud}[s])
-\log_{2}(1+\mathrm{SNR}_{ue}[s]).\label{pro8}
\end{align}
\end{small}%
Similarly, the SNR of the UT intercepting the UAV transmission is given by
\begin{small}
\begin{align}
&\mathrm{SNR}_{ut}[s]=\frac{|L_{urt}[s]\mathbf{h}_{rt}[s]^{H}\boldsymbol{\Theta}[s]\mathbf{H}_{ur}[s]\mathbf{w}[s]|^{2}}{\sigma^{2}}.\label{pro9}
\end{align}
\end{small}%
Finally, based on (\ref{pro6}) and (\ref{pro9}), the communication secrecy rate of of the IoT device with respect to the UT is defined as
\begin{small}
\begin{align}
&\bar{R}_{sec}[s]=\log_{2}(1+\mathrm{SNR}_{ud}[s])
-\log_{2}(1+\mathrm{SNR}_{ut}[s]).\label{pro10}
\end{align}
\end{small}%
Based on \cite{10334744}, for the sensing performance, we derive the SNR of the UT echo signal based on (\ref{pro3}) which results in
\begin{small}
\begin{align}
&\widetilde{\mathrm{SNR}}_{ut}[s]=\frac{|\mathbf{u}[s]^{H}L_{urt}^{2}[s]\mathbf{H}_{ur}[s]^{H}\boldsymbol{\Theta}[s]\mathbf{A}_{rt}\boldsymbol{\Theta}[s]\mathbf{H}_{ur}[s]\mathbf{w}[s]|^{2}}{\mathbf{u}[s]^{H}\mathbf{u}[s]\sigma^{2}}.\label{pro_9}
\end{align}
\end{small}%

\subsection{CSI Uncertainty}
To model the CSI uncertainty of Eve and UT, $\mathrm{SNR}_{ue}[s]$ and $\mathrm{SNR}_{ut}[s]$ can be rewritten as
\begin{small}
\begin{align}
&\mathrm{SNR}_{ue}[s]=\frac{1}{\sigma^{2}}|\mathbf{h}_{e1}^{H}[s]\mathbf{H}_{e1}[s]\mathbf{v}[s]|^{2},\nonumber\\
&\mathrm{SNR}_{ut}[s]=\frac{1}{\sigma^{2}}|\mathbf{h}_{t1}^{H}[s]\mathbf{H}_{t1}[s]\mathbf{v}[s]|^{2},\label{pro12}
\end{align}
\end{small}%
where
\begin{small}
\begin{align}
&\mathbf{v}[s]=\mathbf{H}_{ur}[s]\mathbf{w}[s],
u[s]=\mathbf{h}_{ue}^{H}[s]\mathbf{w}[s],\label{pro13}\\
&\mathbf{H}_{e1}[s]=\mathrm{diag}\left(\left[\begin{matrix}
L_{ure}[s]\mathbf{v}[s]\\
L_{ue}[s]\\
  \end{matrix}
  \right]\right),
  \mathbf{h}_{e1}[s]=\left[\begin{matrix}
\mathbf{h}_{re}[s]\\
u[s]^{H}\\
  \end{matrix}
  \right],\label{pro14}\\
&\mathbf{H}_{t1}[s]=\mathrm{diag}\left(\left[\begin{matrix}
L_{urt}[s]\mathbf{v}[s]\\
  \end{matrix}
  \right]\right),
  \mathbf{h}_{t1}[s]=\left[\begin{matrix}
\mathbf{h}_{rt}[s]\\
  \end{matrix}
  \right],\label{pro15}
\end{align}
\end{small}%
where $\mathbf{H}_{e1}[s]$ and $\mathbf{h}_{e1}[s]$ are the channel of Eve, and $\mathbf{H}_{t1}[s]$ and $\mathbf{h}_{t1}[s]$ are the channel of UT. In general, channel estimation algorithms can be implemented by the UAV to obtain CSI of the IoT device and RIS channels\cite{10334744}. However, Eve typically avoids detecting and tracking legitimate transmitters to intercept legitimate communications. Therefore, there is often uncertainty in Eve's CSI. Based on these considerations, we model the uncertainties of the Eve and UT channel as
\begin{small}
\begin{align}
&\mathbf{h}_{e1}[s]=\bar{\mathbf{h}}_{e1}+\Delta\mathbf{h}_{e1}[s],\Omega_{e1}=\{\Delta\mathbf{h}_{e1}[s]\in\mathbb{C}^{(M+1)\times 1}:\nonumber\\
&\|\Delta\mathbf{h}_{e1}[s]\|\leq\epsilon_{e1},\forall~n\},\label{pro16}\\
&\mathbf{h}_{t1}[s]=\bar{\mathbf{h}}_{t1}+\Delta\mathbf{h}_{t1}[s],\Omega_{t1}=\{\Delta\mathbf{h}_{t1}[s]\in\mathbb{C}^{M\times 1}:\nonumber\\
&\|\Delta\mathbf{h}_{t1}[s]\|\leq\epsilon_{t1},\forall~n\}.\label{pro17}
\end{align}
\end{small}%
Based on (\ref{pro16}) and (\ref{pro17}), the worst-case secrecy rates of Eve and the UT on the $s$-th time slot can be expressed as
\begin{small}
\begin{align}
&R_{sec}^{e1}[s]=\left[R_{ud}[s]-\max_{\Delta\mathbf{h}_{e1}[s]\in\Omega_{e1}}R_{ue}[s]\right]^{+},\nonumber\\
&R_{sec}^{t1}[s]=\left[R_{ud}[s]-\max_{\Delta\mathbf{h}_{t1}[s]\in\Omega_{t1}}R_{ut}[s]\right]^{+},\label{pro19} 
\end{align}
\end{small}%
where $R_{ud}[s]=\log_{2}(1+\mathrm{SNR}_{ud}[s])$, $R_{ue}[s]=\log_{2}(1+\mathrm{SNR}_{ue}[s])$ and $R_{ut}[s]=\log_{2}(1+\mathrm{SNR}_{ut}[s])$.
We aim to optimize the UAV trajectory, RIS phase shift and beamforming based on the weighted average secrecy rate which is given by
\begin{small}
\begin{align}
R_{sum}[s]=\omega R_{sec}^{e1}[s]+(1-\omega)R^{t1}_{sec}[s],\label{pro20}
\end{align}
\end{small}%
where $R_{e1}^{sec}=R_{sec}^{e1}[s]$, $R_{t1}^{sec}=R_{sec}^{t1}[s]$ and $\omega$ is the weight factor which is used to balance the system's secrecy improvement with respect to the Eve and UT in Fig.\ref{FIGURE0}. Therefore, the optimization problem can be expressed as
\begin{small}
\begin{subequations}
\begin{align}
\max_{\bar{\mathbf{p}}_{u}[s],\mathbf{w}[s],\boldsymbol{\Theta}[s],\mathbf{u}[s]}&~\frac{1}{S}\sum_{s=1}^{S}R_{sum}[s],\label{pro21a}\\
\mbox{s.t.}~
&0\leq\theta_{i}[s]<2\pi, i=1,\ldots,N_{R},&\label{pro21b}\\
&\|\bar{\mathbf{p}}_{u}[s+1]-\bar{\mathbf{p}}_{u}[s]\|^{2}\leq D^{2}, s=1,\ldots,S-1,&\label{pro21c}\\
&\|\bar{\mathbf{p}}_{u}[S]-\bar{\mathbf{p}}_{u}[F]\|^{2}\leq D^{2}, \bar{\mathbf{p}}_{u}[1]=\bar{\mathbf{p}}_{u}[0],&\label{pro21d}\\
&\frac{1}{S}\sum_{s=1}^{S}\|\mathbf{w}[s]\|^{2}\leq P,&\label{pro21e}\\
&0\leq \|\mathbf{w}[s]\|^{2}\leq P_{peak},&\label{pro21f}\\
&\widetilde{\mathrm{SNR}}_{ut}[s]\geq \gamma,&\label{pro21g}\\
&\|\mathbf{u}[s]\|_{2}=1.&\label{pro21h}
\end{align}\label{pro21}%
\end{subequations}
\end{small}%
(\ref{pro21b}) denotes the constraint of RIS phase shift. Constraint conditions (\ref{pro21c}) and (\ref{pro21d}) are the mobility constraints of the UAV, where $\bar{\mathbf{p}}_{u}[0]$ and $\bar{\mathbf{p}}_{u}[F]$ are the UAV's initial and final position. $D=v_{max}\Delta_{s}$ denotes the UAV's distance in $\Delta_{s}$, where $v_{max}$ is the maximum speed of the UAV. Constraint condition (\ref{pro21e}) is the transmitted average time power constraint of UAV. Constraint (\ref{pro21f}) is the transmitted power constraint of the UAV on the $s$-th slot. Constraint condition (\ref{pro21g}) is the sensing performance requirement with $\gamma$ being the required minimum SNR of the echo signal at the UAV. 
According to (\ref{pro13}), the problem in (\ref{pro21}) can be equivalently transformed as
\begin{small}
\begin{subequations}
\begin{align}
\max_{\bar{\mathbf{p}}_{u}[s],\mathbf{w}[s],\boldsymbol{\Theta}[s],\mathbf{v}[s],u[s]}&~\frac{1}{S}\sum_{s=1}^{S}R_{sum}[s],\label{pro22a}\\
\mbox{s.t.}~
&(\ref{pro21b}),(\ref{pro21c}),(\ref{pro21d}),(\ref{pro21e}),(\ref{pro21f}),(\ref{pro21g}),(\ref{pro21h}),&\label{pro22b}\\
&\mathbf{v}[s]=\mathbf{H}_{ur}[s]\mathbf{w}[s],&\label{pro22c}\\
&u[s]=\mathbf{h}_{ue}^{H}[s]\mathbf{w}[s].&\label{pro22d}
\end{align}\label{pro22}%
\end{subequations}
\end{small}%
Since both the objective function in (\ref{pro22a}) and the constraints in (\ref{pro22b})are non-convex, the problem in (\ref{pro22}) is non-convex. To solve the above non-convex problem, we propose a BCD-based algorithm to solve the above problem. Next, we introduce in detail how to solve this non-convex problem by alternately optimizing the transmit beamforming vector, UAV trajectory, and receive beamforming vector in the objective function (\ref{pro22}).

\section{Proposed Algorithm for Secrecy Rate Optimization}\label{III}
In this section, to cope with problem (\ref{pro22}), an SCA-based
BCD algorithm is proposed. 
problem (\ref{pro22}) into four subproblems:\\
\subsection{RIS-aided ISAC-UAV Transmitted Beamforming}
We first optimize the ISAC transmit beamforming for a fixed UAV trajectory, RIS phase shift and receive beamforming vector. As such, given $\bar{\mathbf{p}}_{u}[s]$, $\boldsymbol{\Theta}[s]$ and $\mathbf{w}[s]$, the optimization in (\ref{pro22}) simplifies to
\begin{small}
\begin{subequations}
\begin{align}
\max_{\mathbf{w}[s],\mathbf{v}[s],u[s]}&~\frac{1}{S}\sum_{s=1}^{S}R_{sum}[s],\label{pro23a}\\
\mbox{s.t.}~
&(\ref{pro21e}),(\ref{pro21f}),(\ref{pro21g}),(\ref{pro22c}),(\ref{pro22d}),&\label{pro23b}
\end{align}\label{pro23}%
\end{subequations}
\end{small}%
where
\begin{small}
\begin{align}
&R_{sum}[s]=\omega\log_{2}\left(1+\frac{1}{\sigma^{2}}\left|\mathbf{h}_{d1}^{H}[s]\mathbf{H}_{d1}[s]\mathbf{v}[s]\right|^{2}\right)-\nonumber\\
&\omega\log_{2}\left(1+\max_{\Delta\mathbf{h}_{e1}[s]\in\Omega_{e1}}\frac{1}{\sigma^{2}}|\mathbf{h}_{e1}^{H}[s]\mathbf{H}_{e1}[s]\mathbf{v}[s]|^{2}|^{2}\right)\nonumber\\
&+(1-\omega)\left(\log_{2}\left(1+\frac{1}{\sigma^{2}}\left|\mathbf{h}_{d1}^{H}[s]\mathbf{H}_{d1}[s]\mathbf{v}[s]\right|^{2}\right)-\right.\nonumber\\
&\left.\log_{2}\left(1+\max_{\Delta\mathbf{h}_{t1}[s]\in\Omega_{t1}}\frac{1}{\sigma^{2}}\left|\mathbf{h}_{t1}^{H}[s]\mathbf{H}_{t1}[s]\mathbf{v}[s]\right|^{2}|^{2}\right)\right),\nonumber\\
&\mathbf{H}_{d1}[s]=\mathrm{diag}\left(\left[\begin{matrix}
L_{urd}[s]\mathbf{v}[s]\\
L_{ud}[s]\\
  \end{matrix}
  \right]\right),
  \mathbf{h}_{d1}[s]=\left[\begin{matrix}
\mathbf{h}_{rd}[s]\\
\tilde{u}^{H}[s]\\
  \end{matrix}
  \right],\nonumber\\
&\tilde{u}[s]=\mathbf{h}_{ud}^{H}[s]\mathbf{w}[s].\label{pro25}
\end{align}
\end{small}%
According to (\ref{pro25}), problem (\ref{pro23}) is rewritten as
\begin{small}
\begin{subequations}
\begin{align}
\max_{\mathbf{w}[s],\mathbf{v}[s],u[s],\tilde{u}[s]}&~\frac{1}{S}\sum_{s=1}^{S}R_{sum}[s],\label{pro26a}\\
\mbox{s.t.}~
&(\ref{pro21e}),(\ref{pro21f}),(\ref{pro21g}),(\ref{pro22c}),(\ref{pro22d}),&\label{pro26b}\\
&\tilde{u}[s]=\mathbf{h}_{ud}^{H}[s]\mathbf{w}[s].&\label{pro26c}
\end{align}\label{pro26}%
\end{subequations}
\end{small}%
To solve the problem in (\ref{pro26}), we need first to solve the optimization problems in the formula (\ref{pro26a}). The optimization can be equivalently reformulated as
\begin{small}
\begin{align}
\max_{\Delta\mathbf{h}_{e1}[s]\in\Omega_{e1},\Delta\mathbf{h}_{t1}[s]\in\Omega_{t1}}&~|\Delta\mathbf{h}_{e1}^{H}[s]\mathbf{H}_{e1}[s]\mathbf{v}[s]|^{2}+\nonumber\\
&|\Delta\mathbf{h}_{t1}^{H}[s]\mathbf{H}_{t1}[s]\mathbf{v}[s]|^{2}.\label{pro27}
\end{align}
\end{small}%
However, the infinite number of possible CSI uncertainties in $\Delta\mathbf{h}_{e1}[s]$ and $\Delta\mathbf{h}_{t1}[s]$ makes problem (\ref{pro27}) intractable.

According to \cite{boyd2004convex}, this problem can be solved using the unique structure of $\Delta\mathbf{h}_{e1}[s]$ and $\Delta\mathbf{h}_{t1}[s]$. According to the triangle inequality, we have
\begin{small}
\begin{align}
&|\mathbf{h}_{e1}^{H}[s]\mathbf{H}_{e1}[s]\mathbf{v}[s]|\leq |\bar{\mathbf{h}}_{e1}^{H}[s]\mathbf{H}_{e1}[s]\mathbf{v}[s]|+|\Delta\mathbf{h}_{e1}^{H}[s]\mathbf{H}_{e1}[s]\mathbf{v}[s]|,\label{pro28}\\
&|\mathbf{h}_{t1}^{H}[s]\mathbf{H}_{t1}[s]\mathbf{v}[s]|\leq |\bar{\mathbf{h}}_{t1}^{H}[s]\mathbf{H}_{t1}[s]\mathbf{v}[s]|+|\Delta\mathbf{h}_{t1}^{H}[s]\mathbf{H}_{t1}[s]\mathbf{v}[s]|.\label{pro29}
\end{align}
\end{small}%
To make the equal signs in formulas (\ref{pro28}) and (\ref{pro29}) hold, we have
\begin{small}
\begin{align}
&\arg\left(\bar{\mathbf{h}}_{e1}^{H}[s]\mathbf{H}_{e1}[s]\mathbf{v}[s]\right)=\arg\left(\Delta\mathbf{h}_{e1}^{H}[s]\mathbf{H}_{e1}[s]\mathbf{v}[s]\right),\label{pro30}\\
&\arg\left(\bar{\mathbf{h}}_{t1}^{H}[s]\mathbf{H}_{t1}[s]\mathbf{v}[s]\right)=\arg\left(\Delta\mathbf{h}_{t1}^{H}[s]\mathbf{H}_{t1}[s]\mathbf{v}[s]\right).\label{pro31}
\end{align}
\end{small}%
Therefore, problem (\ref{pro27}) can be rewritten
\begin{small}
\begin{subequations}
\begin{align}
\max_{\Delta\mathbf{h}_{e1}[s],\Delta\mathbf{h}_{t1}[s]}&~|\Delta\mathbf{h}_{e1}^{H}[s]\mathbf{H}_{e1}[s]\mathbf{v}[s]|^{2}+|\Delta\mathbf{h}_{t1}^{H}[s]\mathbf{H}_{t1}[s]\mathbf{v}[s]|^{2},\label{pro32a}\\
\mbox{s.t.}~
&(\ref{pro30}),(\ref{pro31}).&\label{pro32b}
\end{align}\label{pro32}%
\end{subequations}
\end{small}%
The objective functions in (\ref{pro32a}) can be rewritten as
\begin{small}
\begin{align}
&\Delta\mathbf{h}_{e1}^{H}[s]\mathbf{H}_{e1}[s]\mathbf{v}[s]+\Delta\mathbf{h}_{t1}^{H}[s]\mathbf{H}_{t1}[s]\mathbf{v}[s]=\Delta\mathbf{h}_{e1}^{H}[s]\boldsymbol{\upsilon}^{e}[s]+\nonumber\\
&\Delta\mathbf{h}_{t1}^{H}[s]\boldsymbol{\upsilon}^{t}[s]=\left(|\Delta h_{e1,1}[s]||c^{e}_{1}[s]|e^{j(\psi_{1}^{e}[s]-\tau_{1}^{e}[s])}+\ldots
+\right.\nonumber\\
&\left.|\Delta h_{e1,M+1}[s]||c^{e}_{M+1}[s]|e^{\psi_{M+1}^{e}[s]-\tau_{M+1}^{e}[s]}\right)+\left(|\Delta h_{t1,1}[s]||c^{t}_{1}[s]|\right.\nonumber\\
&\left.e^{j(\psi_{1}^{t}[s]-\tau_{1}^{t}[s])}+\ldots+|\Delta h_{t1,M}[s]||c^{t}_{M}[s]|e^{\psi_{M}^{t}[s]-\tau_{M}^{t}[s]}\right),\label{pro33}
\end{align}
\end{small}%
where $\Delta\mathbf{h}_{e1}[s]=\left[|\Delta h_{e1,1}[s]|e^{j\tau_{1}[s]},|\Delta h_{e1,2}[s]|e^{j\tau_{2}[s]},\ldots,\right.$ 
$\left.
|\Delta h_{e1,M+1}[s]|e^{j\tau_{M+1}[s]}\right]$,
$\boldsymbol{\upsilon}^{e}[s]=\mathbf{H}_{e1}[s]$ $\mathbf{v}[s]=\left[|c_{1}^{e}[s]|e^{j\psi_{1}[s]},|c_{2}^{e}[s]|e^{j\psi_{2}[s]},\ldots,|c^{e}_{M+1}[s]|e^{j\psi_{M+1}[s]}\right]$, $\Delta\mathbf{h}_{t1}[s]$\\
$=\left[|\Delta h_{t1,1}[s]|e^{j\tau_{1}[s]},|\Delta h_{t1,2}[s]|\right.$ 
$\left.e^{j\tau_{2}},\ldots,|\Delta h_{t1,M}[s]|\right.$ $\left.e^{j\tau_{M}[s]}\right]$,
$\boldsymbol{\upsilon}^{t}[s]=\mathbf{H}_{t1}[s]\mathbf{v}[s]=\left[|c_{1}^{t}[s]|e^{j\psi_{1}^{t}[s]},|c_{2}^{t}[s]|e^{j\psi_{2}^{t}[s]}\right.$ 
$\left.,\ldots,|c^{t}_{M}[s]|e^{j\psi_{M}[s]}\right]$.
According to \cite{8811733}, to maximize the objective functions in (\ref{pro32a}), we have $\psi_{1}^{e}[s]-\tau_{1}^{e}[s]=\ldots=\psi_{M+1}^{e}[s]-\tau_{M+1}^{e}[s]$ and  $\psi_{1}^{t}[s]-\tau_{1}^{t}[s]=\ldots=\psi_{M}^{t}[s]-\tau_{M}^{t}[s]$. Based on constraint conditions in (\ref{pro30}) and (\ref{pro31}), we have
\begin{small}
\begin{align}
&\tau_{k,e1}^{op}[s]=\psi_{k}^{e}[s]-\arg(\bar{\mathbf{h}}_{e1}^{H}\mathbf{H}_{e1}[s]\mathbf{v}[s]),\nonumber\\
&\tau_{k,t1}^{op}[s]=\psi_{k}^{t}[s]-\arg(\bar{\mathbf{h}}_{t1}^{H}\mathbf{H}_{t1}[s]\mathbf{v}[s]).\label{pro35}
\end{align}
\end{small}%
Based on (\ref{pro33}) and (\ref{pro35}), problem (\ref{pro32}) can be rewritten as
\begin{subequations}
\begin{align}
\max_{\boldsymbol{\nu}_{1}^{e}[s],\boldsymbol{\nu}_{1}^{t}[s]}&~|(\boldsymbol{\nu}_{1}^{e})^{T}[s]\boldsymbol{\nu}_{2}^{e}[s]|^{2}+|(\boldsymbol{\nu}_{1}^{t})^{T}[s]\boldsymbol{\nu}_{2}^{t}[s]|^{2},\label{pro36a}\\
\mbox{s.t.}~
&\|\boldsymbol{\nu}_{1}^{e}[s]\|\leq\epsilon_{1},&\label{pro36b}\\
&\|\boldsymbol{\nu}_{1}^{t}[s]\|\leq\epsilon_{2},&\label{pro36c}
\end{align}\label{pro36}%
\end{subequations}
where
$\boldsymbol{\nu}_{1}^{e}[s]=[|\Delta h_{e1,1}[s]|,|\Delta h_{e1,2}[s]|,\ldots,|\Delta h_{e1,M+1}[s]|]^{T}$,\\
$
\boldsymbol{\nu}_{2}^{e}[s]=[|c_{1}^{e}[s]|,|c_{2}^{e}[s]|,\ldots,|c_{M+1}^{e}[s]|]^{T}$, $\boldsymbol{\nu}_{1}^{t}[s]=[|\Delta h_{t1,1}[s]|,|\Delta h_{t1,2}[s]|,\ldots,|\Delta h_{t1,M}[s]|]^{T}$,$
\boldsymbol{\nu}_{2}^{t}[s]=[|c_{1}^{t}[s]|,|c_{2}^{t}[s]|,\ldots,|c_{M}^{t}[s]|]^{T}$.
the optimization solutions of problem (\ref{pro36}) $
\tilde{\boldsymbol{\nu}}_{1}^{e}[s]=\frac{\epsilon_{1}}{\|\boldsymbol{\nu}_{2}^{e}[s]\|}\boldsymbol{\nu}_{2}^{e}[s]$ and $
\tilde{\boldsymbol{\nu}}_{1}^{t}[s]=\frac{\epsilon_{2}}{\|\boldsymbol{\nu}_{2}^{t}[s]\|}\boldsymbol{\nu}_{2}^{t}[s]$, respectively. Furthermore, the optimal solution to problem (\ref{pro36}) is expressed as
\begin{small}
\begin{align}
&\widetilde{\Delta\mathbf{h}}_{e1}=\mathrm{diag}\left(e^{j\tau_{1}^{e}[s]},e^{j\tau_{2}^{e}[s]},\ldots,e^{j\tau_{M+1}^{e}[s]}\right)\tilde{\boldsymbol{\nu}}_{1}^{e}[s],\nonumber\\
&\widetilde{\Delta\mathbf{h}}_{t1}=\mathrm{diag}\left(e^{j\tau_{1}^{t}[s]},e^{j\tau_{2}^{t}[s]},\ldots,e^{j\tau_{M}^{t}[s]}\right)\tilde{\boldsymbol{\nu}}_{1}^{t}[s].\label{pro37}
\end{align}    
\end{small}%
Based on (\ref{pro37}), problem (\ref{pro26}) can be rewritten as
\begin{small}
\begin{subequations}
\begin{align}
\max_{\mathbf{w}[s],\mathbf{v}[s],u[s],\bar{u}[s],\tilde{u}[s]}&~\frac{1}{T}\sum_{t=1}^{T}\widetilde{R_{sum}}[s],\label{pro38a}\\
\mbox{s.t.}~
&(\ref{pro21e}),(\ref{pro21f}),(\ref{pro21g}),(\ref{pro22c}),(\ref{pro22d}),(\ref{pro26c}),&\label{pro38b}
\end{align}\label{pro38}%
\end{subequations}
\end{small}%
where 
\begin{small}
\begin{align}
&\widetilde{R_{sum}}[s]=\omega\log_{2}\left(1+\frac{1}{\sigma^{2}}\left|\mathbf{h}_{d1}^{H}[s]\mathbf{H}_{d1}[s]\mathbf{v}[s]\right|^{2}\right)-\nonumber\\
&\omega\log_{2}\left(1+\frac{1}{\sigma^{2}}|\widetilde{\mathbf{h}_{e1}}^{H}[s]\mathbf{H}_{e1}[s]\mathbf{v}[s]|^{2}\right)\nonumber\\
&+(1-\omega)\left(\log_{2}\left(1+\frac{1}{\sigma^{2}}\left|\mathbf{h}_{d1}^{H}[s]\mathbf{H}_{d1}[s]\mathbf{v}[s]\right|^{2}\right)-\right.\nonumber\\
&\left.\log_{2}\left(1+\frac{1}{\sigma^{2}}\left|\widetilde{\mathbf{h}_{t1}}^{H}[s]\mathbf{H}_{t1}[s]\mathbf{v}[s]\right|^{2}\right)\right),\nonumber\\
&\widetilde{\mathbf{h}_{e1}}[s]=\mathbf{h}_{e1}[s]+\widetilde{\Delta\mathbf{h}}_{e1}[s],~\widetilde{\mathbf{h}_{t1}}[s]=\mathbf{h}_{t1}[s]+\widetilde{\Delta\mathbf{h}}_{t1}[s].\label{pro39}%
\end{align}
\end{small}%
Given that the problem in (\ref{pro39}) is non-convex, we use SCA to solve problem (\ref{pro39}) by introducing slack variable $\zeta_{1}[s]=|\widetilde{\mathbf{h}_{e1}}^{H}[s]\mathbf{H}_{e1}[s]\mathbf{v}[s]|^{2}$, $\zeta_{2}[s]=|\widetilde{\mathbf{h}_{t1}}^{H}[s]\mathbf{H}_{t1}[s]\mathbf{v}[s]|^{2}$. According to Taylor series, the upper bound of $|\mathbf{u}[s]^{H}L_{urt}^{2}[s]\mathbf{H}_{ur}[s]^{H}\boldsymbol{\Theta}[s]\mathbf{A}_{rt}\boldsymbol{\Theta}[s]\mathbf{H}_{ur}[s]\mathbf{w}[s]|^{2}$ is denoted as
\begin{small}
\begin{align}
&|\mathbf{u}[s]^{H}L_{urt}^{2}[s]\mathbf{H}_{ur}[s]^{H}\boldsymbol{\Theta}[s]\mathbf{A}_{rt}\boldsymbol{\Theta}[s]\mathbf{H}_{ur}[s]\mathbf{w}[s]|^{2}\nonumber\\
&\leq|\mathbf{u}[s]^{H}L_{urt}^{2}[s]\mathbf{H}_{ur}[s]^{H}\boldsymbol{\Theta}[s]\mathbf{A}_{rt}\boldsymbol{\Theta}[s]\mathbf{H}_{ur}[s]\mathbf{w}_{0}[s]|^{2}+\mathbf{u}[s]^{H}\nonumber\\
&L_{urt}^{2}[s]\mathbf{H}_{ur}[s]^{H}\boldsymbol{\Theta}[s]\mathbf{A}_{rt}\boldsymbol{\Theta}[s]\mathbf{H}_{ur}[s]\mathbf{w}_{0}[s](\mathbf{w}[s]-\mathbf{w}_{0}[s]).\label{pro40}
\end{align}
\end{small}%
Therefore, the constraint condition in (\ref{pro21g}) is expressed as
\begin{small}
\begin{align}
&\mathrm{SNR}_{ut}[s]\leq|\mathbf{u}[s]^{H}L_{urt}^{2}[s]\mathbf{H}_{ur}[s]^{H}\boldsymbol{\Theta}[s]\mathbf{A}_{rt}\boldsymbol{\Theta}[s]\mathbf{H}_{ur}[s]\mathbf{w}_{0}[s]|^{2}\nonumber\\
&+\mathbf{u}[s]^{H}L_{urt}^{2}[s]\mathbf{H}_{ur}[s]^{H}\boldsymbol{\Theta}[s]\mathbf{A}_{rt}\boldsymbol{\Theta}[s]\mathbf{H}_{ur}[s]\mathbf{w}_{0}[s](\mathbf{w}[s]-\nonumber\\
&\mathbf{w}_{0}[s]).\label{pro41}
\end{align}
\end{small}%
We also use the approximation of the first order Taylor series of $\widetilde{R_{sum}}[s]$ which is denoted as
\begin{small}
\begin{align}
&\widetilde{R_{sum}}[s]=\omega\log_{2}\left(1+\frac{1}{\sigma^{2}}\left|\mathbf{h}_{d1}^{H}[s]\mathbf{H}_{d1}[s]\mathbf{v}[s]\right|^{2}\right)\nonumber\\
&+(1-\omega)\log_{2}\left(1+\frac{1}{\sigma^{2}}\left|\mathbf{h}_{d1}^{H}[s]\mathbf{H}_{d1}[s]\mathbf{v}[s]\right|^{2}\right)\nonumber\\
&-\omega\log_{2}\left(1+\frac{\zeta_{1}[s]}{\sigma^{2}}\right)-(1-\omega)\log_{2}\left(1+\frac{\zeta_{2}[s]}{\sigma^{2}}\right)\nonumber\\
&\leq \omega\log_{2}\left(1+\frac{1}{\sigma^{2}}\left|\mathbf{h}_{d1}^{H}[s]\mathbf{H}_{d1}[s]\mathbf{v}[s]\right|^{2}\right)+\nonumber\\
&(1-\omega)\log_{2}\left(1+\frac{1}{\sigma^{2}}\left|\mathbf{h}_{d1}^{H}[s]\mathbf{H}_{d1}[s]\mathbf{v}[s]\right|^{2}\right)-\omega\log_{2}\left(1+\frac{\zeta_{1}[s]}{\sigma^{2}}\right)\nonumber\\
&-\omega\frac{1}{\ln2\left(1+\frac{\zeta_{1,0}[s]}{\sigma^{2}}\right)}(\zeta_{1}[s]-\zeta_{1,0}[s])-(1-\omega)\log_{2}\left(1+\frac{\zeta_{2,0}[s]}{\sigma^{2}}\right)\nonumber\\
&-(1-\omega)
\frac{1}{\ln2\left(1+\frac{\zeta_{2,0}[s]}{\sigma^{2}}\right)}(\zeta_{2}[s]-\zeta_{2,0}[s])=\widehat{R_{sum}}[s].\label{pro42}
\end{align}
\end{small}%
Problem (\ref{pro38}) is expressed as
\begin{small}
\begin{subequations}
\begin{align}
\max_{\mathbf{w}[s],\mathbf{v}[s],u[s],\bar{u}[s],\tilde{u}[s]}&~\frac{1}{S}\sum_{s=1}^{S}\widehat{R_{sum}}[s],\label{pro43a}\\
\mbox{s.t.}~
&(\ref{pro21e}),(\ref{pro21f}),(\ref{pro22c}),(\ref{pro22d}),(\ref{pro26c}),(\ref{pro41}),&\label{pro43b}\\
&|\widetilde{\mathbf{h}_{e1}}^{H}[s]\mathbf{H}_{e1}[s]\mathbf{v}[s]|^{2}\leq\zeta_{1}[s],\nonumber\\
&|\widetilde{\mathbf{h}_{t1}}^{H}[s]\mathbf{H}_{t1}[s]\mathbf{v}[s]|^{2}\leq\zeta_{2}[s].&\label{pro43c}
\end{align}\label{pro43}%
\end{subequations}
\end{small}%
According to \cite{boyd2004convex}, constraint conditions and objective function in (\ref{pro21e}),(\ref{pro21f}),(\ref{pro22c}),(\ref{pro22d}),(\ref{pro26c}),(\ref{pro41}), (\ref{pro43b}) and (\ref{pro43a}) are convex. Finally, we can use CVX in \cite{grant2009cvx}  to solve problem (\ref{pro43}).

\subsection{RIS-aided ISAC-UAV RIS Phase Shift Matrix}
Given $\bar{\mathbf{p}}_{u}[s]$, $\mathbf{w}[s]$ and $\mathbf{u}[s]$, we have
\begin{small}
\begin{subequations}
\begin{align}
\max_{\boldsymbol{\Theta}[s]}&~\frac{1}{S}\sum_{s=1}^{S}\widetilde{R_{sum}}[s],\label{pro44a}\\
\mbox{s.t.}~
&0\leq\theta_{i}[s]<2\pi,&\label{pro44b}\\
&\frac{|\mathbf{u}[s]^{H}L_{urt}^{2}[s]\mathbf{H}_{ur}[s]^{H}\boldsymbol{\Theta}[s]\mathbf{A}_{rt}\boldsymbol{\Theta}[s]\mathbf{H}_{ur}[s]\mathbf{w}[s]|^{2}}{\mathbf{u}[s]^{H}\mathbf{u}[s]\sigma^{2}}
\geq \gamma_{T}.&\label{pro44c}
\end{align}\label{pro44}%
\end{subequations}
\end{small}%
To facilitate the analysis, we introduce auxiliary variables $\mathbf
{F}_{U}[s]$,$\mathbf
{F}_{E}[s]$,$\mathbf
{F}_{T}[s]$,$\mathbf
{S}_{1}[s]$ and $\mathbf{S}_{2}[s]$  by defining $\boldsymbol{\Theta}[s]=\mathrm{diag}(\boldsymbol{\theta}[s])$, $\tilde{\mathbf{Q}}[s]=\tilde{\boldsymbol{\theta}}[s]\tilde{\boldsymbol{\theta}}^{H}[s]$, $\tilde{\boldsymbol{\theta}}[s]=[1,\boldsymbol{\theta}^{T}[s]]^{T}$ and
\begin{small}
\begin{align}
&\mathbf{F}_{U}[s]=L_{ud}[s]\mathbf{h}_{ud}^{H}[s]\mathbf{w}[s]\mathbf{w}^{H}[s]\mathbf{h}_{ud}[s]+\mathrm{diag}\{\mathbf{w}^{H}[s]\mathbf{H}_{ur}^{H}[s]\}\nonumber\\
&\mathbf{h}_{rd}\mathbf{h}_{rd}^{H}\mathrm{diag}\{\mathbf{H}_{ur}[s]\mathbf{w}[s]\},\nonumber\\
&\mathbf{F}_{E}[s]=L_{ue}[s]\mathbf{h}_{ue}^{H}[s]\mathbf{w}[s]\mathbf{w}^{H}[s]\mathbf{h}_{ue}[s]+\mathrm{diag}\{\mathbf{w}^{H}[s]\mathbf{H}_{ur}^{H}[s]\}\nonumber\\
&\mathbf{h}_{re}\mathbf{h}_{re}^{H}\mathrm{diag}\{\mathbf{H}_{ur}[s]\mathbf{w}[s]\},\nonumber\\
&\mathbf{F}_{T}[s]=\mathrm{diag}\{\mathbf{w}^{H}[s]\mathbf{H}_{ur}^{H}[s]\}\mathbf{h}_{rt}\mathbf{h}_{rt}^{H}\mathrm{diag}\{\mathbf{H}_{ur}[s]\mathbf{w}[s]\},\nonumber\\
&\mathbf{C}_{1}[s]=L_{urt}^{2}\mathbf{H}_{ur}[s]\mathbf{u}[s]\mathbf{u}^{H}[s]\mathbf{H}_{ur}^{H}[s],\nonumber\\
&\mathbf{C}_{2}[s]=\mathrm{diag}\{\mathbf{w}^{H}[s]\mathbf{H}^{H}_{ur}[s]\}\mathbf{A}_{rt}\mathrm{diag}\{\mathbf{H}_{ur}[s]\mathbf{w}^{H}[s]\}.\label{pro45}
\end{align}
\end{small}%
Problem (\ref{pro44}) is expressed as
\begin{small}
\begin{subequations}
\begin{align}
\max_{\tilde{\mathbf{Q}}[s]\succeq \mathbf{0}}&~\log_{2}\left(1+\frac{\mathrm{tr}(\mathbf{F}_{U}[s]\tilde{\mathbf{Q}}[s])}{\mathrm{tr}(\mathbf{F}_{E}[s]\tilde{\mathbf{Q}}[s])+\sigma^{2}_{u}}\right)\nonumber\\
&-\log_{2}\left(1+\frac{\mathrm{tr}(\mathbf{F}_{U}[s]\tilde{\mathbf{Q}}[s])}{\mathrm{tr}(\mathbf{F}_{T}[s]\tilde{\mathbf{Q}}[s])+\sigma^{2}_{e}}\right),\label{pro46a}\\
\mbox{s.t.}~
&\tilde{\mathbf{Q}}[s]\succeq\mathbf{0},&\label{pro46b}\\
&\mathrm{tr}\left(\mathbf{B}_{i}\tilde{\mathbf{Q}}[s]\right)=1,&\label{pro46c}\\
&\mathrm{tr}\left(\mathbf{C}_{1}[s]\tilde{\mathbf{Q}}[s]\mathbf{C}_{2}[s]\tilde{\mathbf{Q}}[s]\right)\geq \gamma_{s}\sigma_{b}^{2}\mathbf{u}^{H}\mathbf{u},&\label{pro46d}\\
&\mathrm{rank}\left(\tilde{\mathbf{Q}}[s]\right)=1,&\label{pro46e}
\end{align}\label{pro46}%
\end{subequations}
\end{small}%
where $\mathbf{B}_{i}$ is a selection matrix, whose elements are given by
\begin{small}
\begin{align}
[\mathbf{B}_{i}]_{(m,n)}=\left\{
\begin{aligned}
1,&~m=n=i\\
0,&~\mathrm{otherwise}
\end{aligned}
\right..
\end{align}
\end{small}%
Constraint conditions (\ref{pro46b}) and (\ref{pro46c}) are convex. The objective function in (\ref{pro46a}) is rewritten as
\begin{small}
\begin{align}
&f(\tilde{\mathbf{Q}}[s])=\log_{2}\left(\mathrm{tr}((\mathbf{F}_{U}[s]+\mathbf{F}_{E}[s])\tilde{\mathbf{Q}}[s])+\sigma_{u}^{2}\right)-\log_{2}\left(\mathrm{tr}(\mathbf{F}_{E}[s]\right.\nonumber\\
&\left.\tilde{\mathbf{Q}}[s])+\sigma_{u}^{2}\right)-\log_{2}\left(\mathrm{tr}((\mathbf{F}_{U}[s]+\mathbf{F}_{T}[s])\tilde{\mathbf{Q}}[s])+\sigma_{u}^{2}\right)+\log_{2}\nonumber\\
&\left(\mathrm{tr}(\mathbf{F}_{T}[s]\tilde{\mathbf{Q}}[s])+\sigma_{u}^{2}\right)=\log_{2}\left(\mathrm{tr}((\mathbf{F}_{U}[s]+\mathbf{F}_{E}[s])\tilde{\mathbf{Q}}[s])+\sigma_{u}^{2}\right)\nonumber\\
&+\log_{2}\left(\mathrm{tr}(\mathbf{F}_{T}[s]\tilde{\mathbf{Q}}[s])+\sigma_{u}^{2}\right)-\left(\log_{2}\left(\mathrm{tr}(\mathbf{F}_{E}[s]\tilde{\mathbf{Q}}[s])+\sigma_{u}^{2}\right)+\right.\nonumber\\
&\left.\log_{2}\left(\mathrm{tr}((\mathbf{F}_{U}[s]+\mathbf{F}_{T}[s])\tilde{\mathbf{Q}}[s])+\sigma_{u}^{2}\right)\right).\label{pro48}
\end{align}
\end{small}%
According to SCA algorithm in \cite{9472958}, the lower bound of $f(\tilde{\mathbf{Q}}[s])$ can be expressed as
\begin{small}
\begin{align}
&h(\tilde{\mathbf{Q}}[s])
=\log_{2}\left(\mathrm{tr}((\mathbf{F}_{U}[s]+\mathbf{F}_{E}[s])\tilde{\mathbf{Q}}[s])+\sigma_{u}^{2}\right)+\log_{2}\left(\mathrm{tr}(\mathbf{F}_{T}[s]\right.\nonumber\\
&\left.\tilde{\mathbf{Q}}[s])+\sigma_{u}^{2}\right)-\left(\log_{2}\left(\mathrm{tr}(\mathbf{F}_{E}[s]\tilde{\mathbf{Q}}_{0}[s])+\sigma_{u}^{2}\right)+\log_{2}\left(\mathrm{tr}((\mathbf{F}_{U}[s]\right.\right.\nonumber\\
&\left.\left.+\mathbf{F}_{T}[s])\tilde{\mathbf{Q}}_{0}[s])+\sigma_{u}^{2}\right)\right)-\frac{1}{\ln 2(\mathrm{tr}(\mathbf{F}_{E}[s]\tilde{\mathbf{Q}}_{0}[s])+\sigma_{u}^{2})}\mathrm{tr}(\mathbf{F}_{E}[s](\tilde{\mathbf{Q}}[s]\nonumber\\
&-\tilde{\mathbf{Q}}_{0}[s]))-\frac{1}{\ln 2(\mathrm{tr}((\mathbf{F}_{T}[s]+\mathbf{F}_{U}[s])\tilde{\mathbf{Q}}_{0}[s])+\sigma_{u}^{2})}\mathrm{tr}((\mathbf{F}_{T}[s]+\nonumber\\
&\mathbf{F}_{U}[s])(\tilde{\mathbf{Q}}[s]-\tilde{\mathbf{Q}}_{0}[s])).\label{pro49}
\end{align}
\end{small}%
Thus, we have $f(\tilde{\mathbf{Q}}[s])\leq h(\tilde{\mathbf{Q}}[s])$, problem (\ref{pro46}) is written as
\begin{small}
\begin{subequations}
\begin{align}
\max_{\tilde{\mathbf{Q}}[s]\succeq \mathbf{0}}&~h(\tilde{\mathbf{Q}}[s]),\label{pro50a}\\
\mbox{s.t.}~
&(\ref{pro46b}), (\ref{pro46c}),(\ref{pro46d}), (\ref{pro46e}),&\label{pro50b}
\end{align}\label{pro50}%
\end{subequations}
\end{small}%
The objective function in (\ref{pro50a}) and the constraint conditions in (\ref{pro46b}) and (\ref{pro46c}) are convex in the problem (\ref{pro50}). Next, to deal with the non-convex (\ref{pro46d}), we aim to derive a lower bound of $\mathrm{tr}(\mathbf{C}_{1}[s]\tilde{\mathbf{Q}}[s]\mathbf{C}_{2}[s]\tilde{\mathbf{Q}}[s])$, which is further used to transform (\ref{pro46d}) into a convex constraint. Specifically, 
$\mathrm{tr}(\mathbf{C}_{1}[s]\tilde{\mathbf{Q}}[s]\mathbf{C}_{2}[s]\tilde{\mathbf{Q}}[s])$ is lower-bounded by
\begin{small}
\begin{align}
&\mathrm{tr}(\mathbf{C}_{2}[s]\tilde{\mathbf{Q}}[s]\mathbf{C}_{2}[s]\tilde{\mathbf{Q}}[s])\geq\left\|\mathrm{tr}\left(\mathbf{U}\bar{\boldsymbol{\Xi}}^{1/2}\tilde{\mathbf{Q}}_{0}[s]\right)\right\|_{2}^{-2}\nonumber\\
&\mathrm{tr}\left(\mathbf{U}\bar{\boldsymbol{\Xi}}^{1/2}\tilde{\mathbf{Q}}_{0}[s]\mathbf{U}\bar{\boldsymbol{\Xi}}^{1/2}\tilde{\mathbf{Q}}[s]\right)^{2},\label{pro51}
\end{align}
\end{small}%
where $\tilde{\mathbf{Q}}_{0}[s]\succeq\mathbf{0}$ represents a certain point of variable $\tilde{\mathbf{Q}}[s]$. The proof is detailed in \textbf{Appendix}~\ref{appA}. Therefore, constraint condition (\ref{pro46d}) is written as
\begin{small}
\begin{align}
&\left\|\mathrm{tr}\left(\mathbf{U}\bar{\boldsymbol{\Xi}}^{1/2}\tilde{\mathbf{Q}}_{0}[s]\right)\right\|_{2}^{-2}\mathrm{tr}\left(\mathbf{U}\bar{\boldsymbol{\Xi}}^{1/2}\tilde{\mathbf{Q}}_{0}[s]\mathbf{U}\bar{\boldsymbol{\Xi}}^{1/2}\tilde{\mathbf{Q}}[s]\right)^{2}\nonumber\\
&\geq \gamma_{T}\sigma_{b}^{2}\mathbf{u}^{H}\mathbf{u}\nonumber\\
&\Rightarrow\left\|\mathrm{tr}\left(\mathbf{U}\bar{\boldsymbol{\Xi}}^{1/2}\tilde{\mathbf{Q}}_{0}[s]\right)\right\|_{2}^{-1}\mathrm{tr}\left(\mathbf{U}\bar{\boldsymbol{\Xi}}^{1/2}\tilde{\mathbf{Q}}_{0}[s]\mathbf{U}\bar{\boldsymbol{\Xi}}^{1/2}\tilde{\mathbf{Q}}[s]\right)\nonumber\\
&\geq \sqrt{\gamma_{s}\sigma_{b}^{2}}.\label{pro53}
\end{align}
\end{small}%
Constraint condition (\ref{pro53}) is convex. Problem (\ref{pro50}) is rewritten as
\begin{small}
\begin{subequations}
\begin{align}
\max_{\tilde{\mathbf{Q}}[s]\succeq \mathbf{0}}&~h(\tilde{\mathbf{Q}}[s]),\label{pro54a}\\
\mbox{s.t.}~
&(\ref{pro46b}), (\ref{pro46c}),(\ref{pro53}), (\ref{pro46e}).&\label{pro54b}
\end{align}\label{pro54}%
\end{subequations}
\end{small}%
Except for constraint condition (\ref{pro46e}), the objective function and constraints in the problem (\ref{pro54}) are all convex. Therefore, we can use SDP method to solve problem (\ref{pro54}). Finally, it is noted that although we ignore the constraint condition in (\ref{pro46e}), the optimal solution of (\ref{pro54}) still satisfies the rank one constraint. The proof is given in \textbf{Appendix}~\ref{appB}. After $\tilde{\boldsymbol{\theta}}$ is obtained, $\boldsymbol{\theta}=\frac{\tilde{\boldsymbol{\theta}}_{1:N_{R}}}{\tilde{\boldsymbol{\theta}}{N_{R}+1}}$.

\subsection{RIS-aided ISAC-UAV Trajectory Optimization}
Next, we need to optimise the ISAC-UAV trajectory for given transmit and receive beamforming vectors. Since all LoS channel components are complex and non-linear, which makes the UAV's trajectory optimization problem difficult, we proceed by applying the following approximations. We use the UAV trajectory of iteration $(j-1)$-th to approximate all LoS channel components in the $j$-th iteration to approximate the SNR terms.
Then, we can re-express problem (\ref{pro21}) for a given $\mathbf{w}[s]$, $\mathbf{u}[s]$ and $\boldsymbol{\Theta}[s]$, we have 
\begin{small}
\begin{subequations}
\begin{align}
\max_{\bar{\mathbf{p}}_{u}[s]}&~\frac{1}{S}\sum_{n=1}^{N}R_{sum}[s],\label{pro55a}\\
\mbox{s.t.}~
&(\ref{pro21c}), (\ref{pro21d}),(\ref{pro21g}).&\label{pro55b}
\end{align}\label{pro55}%
\end{subequations}
\end{small}%
According to $(\ref{pro6})$, $|L_{ud}[s]\mathbf{h}_{ud}^{H}[s]\mathbf{w}[s]+L_{urd}[s]\mathbf{h}_{rd}^{H}[s]\boldsymbol{\Theta}[s]\mathbf{H}_{ur}[s]\mathbf{w}[s]|^{2}$ can be rewritten as
\begin{small}
\begin{align}
&|L_{ud}[s]\mathbf{h}_{ud}^{H}[s]\mathbf{w}[s]+L_{urd}[s]\mathbf{h}_{rd}^{H}\boldsymbol{\Theta}[s]\mathbf{H}_{ur}[s]\mathbf{w}[s]|^{2}\nonumber\\
&=|\sqrt{\rho (d_{ud}[s])^{-\kappa}}\mathbf{h}_{ud}^{H}[s]\mathbf{w}[s]+\sqrt{\rho(d_{ur}[s]d_{rd})^{-\alpha}}\mathbf{h}_{rd}^{H}\boldsymbol{\Theta}[s]\nonumber\\
&\mathbf{H}_{ur}[s]\mathbf{w}[s]|^{2}=\left[\sqrt{(d_{ud}[s])^{-\kappa}},\sqrt{(d_{ur}[s])^{-\alpha}}\right]\nonumber\\
&\left[\begin{matrix}
\mathbf{h}_{ud}^{H}[s]\mathbf{w}[s]\\
\mathbf{w}^{H}[s]\mathbf{H}_{ur}^{H}[s]\boldsymbol{\Theta}^{H}[s]\mathbf{h}_{rd}\\
  \end{matrix}
  \right]\left[\begin{matrix}
\mathbf{h}_{ud}^{H}[s]\mathbf{w}[s]\\
\mathbf{w}^{H}[s]\mathbf{H}_{ur}^{H}[s]\boldsymbol{\Theta}^{H}[s]\mathbf{h}_{rd}\\
  \end{matrix}
  \right]^{H}\nonumber\\
&\left[\sqrt{(d_{ud}[s])^{-\kappa}},\sqrt{(d_{ur}[s])^{-\alpha}}\right]^{H}.\label{pro56}
\end{align}
\end{small}%
and
\begin{small}
\begin{align}
&|L_{ue}[s]\mathbf{h}_{ue}^{H}[s]\mathbf{w}[s]+L_{ure}[s]\mathbf{h}_{re}^{H}\boldsymbol{\Theta}[s]\mathbf{H}_{ur}[s]\mathbf{w}[s]|^{2}\nonumber\\
&=\left[\sqrt{(d_{ue}[s])^{-\kappa}},\sqrt{(d_{ur}[s])^{-\alpha}}\right]\left[\begin{matrix}
\mathbf{h}_{ue}^{H}[s]\mathbf{w}[s]\\
\mathbf{w}^{H}[s]\mathbf{H}_{ur}^{H}[s]\boldsymbol{\Theta}^{H}[s]\mathbf{h}_{re}\\
  \end{matrix}
  \right]\nonumber\\
&\left[\begin{matrix}
\mathbf{h}_{ue}^{H}[s]\mathbf{w}[s]\\
\mathbf{w}^{H}[s]\mathbf{H}_{ur}^{H}[s]\boldsymbol{\Theta}^{H}[s]\mathbf{h}_{re}\\
  \end{matrix}
  \right]^{H}\left[\sqrt{(d_{ue}[s])^{-\kappa}},\sqrt{(d_{ur}[s])^{-\alpha}}\right]^{H}.\label{pro57}
\end{align}
\end{small}%
$|L_{urt}[s]\mathbf{h}_{rt}^{H}\boldsymbol{\Theta}[s]\mathbf{H}_{ur}[s]\mathbf{w}[s]|^{2}$ is denoted as
\begin{small}
\begin{align}
&|L_{urt}[s]\mathbf{h}_{rt}^{H}\boldsymbol{\Theta}[s]\mathbf{H}_{ur}[s]\mathbf{w}[s]|^{2}\nonumber\\
&=\left[\sqrt{(d_{ut}[s])^{-\kappa}},\sqrt{(d_{ur}[s])^{-\alpha}}\right]\left[\begin{matrix}
0\\
\mathbf{w}^{H}[s]\mathbf{H}_{ur}^{H}[s]\boldsymbol{\Theta}^{H}[s]\mathbf{h}_{rt}\\
  \end{matrix}
  \right]\nonumber\\
&\left[\begin{matrix}
0\\
\mathbf{w}^{H}[s]\mathbf{H}_{ur}^{H}[s]\boldsymbol{\Theta}^{H}[s]\mathbf{h}_{rt}\\
  \end{matrix}
  \right]^{H}\left[\sqrt{(d_{ut}[s])^{-\kappa}},\sqrt{(d_{ur}[s])^{-\alpha}}\right]^{H}.\label{pro58}
\end{align}
\end{small}%
To simplify this expression, we let $\boldsymbol{\Psi}_{1}=\left[\begin{matrix}
\mathbf{h}_{ue}^{H}\mathbf{w}[s]\\
\mathbf{w}^{H}[s]\mathbf{H}_{ur}^{H}[s]\boldsymbol{\Theta}^{H}[s]\mathbf{h}_{re}\\
  \end{matrix}
  \right]$, $\boldsymbol{\Psi}_{2}=\left[\begin{matrix}
\mathbf{h}_{ue}^{H}[s]\mathbf{w}[s]\\
\mathbf{w}^{H}[s]\mathbf{H}_{ur}^{H}[s]\boldsymbol{\Theta}^{H}[s]\mathbf{h}_{re}\\
  \end{matrix}
  \right]$ and $\boldsymbol{\Psi}_{3}=\left[\begin{matrix}
0\\
\mathbf{w}^{H}[s]\mathbf{H}_{ur}^{H}[s]\boldsymbol{\Theta}^{H}[s]\mathbf{h}_{re}\\
  \end{matrix}
  \right]$ and the SNRs in the communication secrecy rate in (6) can be approximated as
\begin{small}
\begin{align}
&R_{ud}[s]=\log_{2}\left(1+[\sqrt{(d_{ud}[s])^{-\kappa}},\sqrt{(d_{ur}[s])^{-\alpha}}]\boldsymbol{\Psi}_{1}\boldsymbol{\Psi}_{1}^{H}\right.\nonumber\\
&\left.[\sqrt{(d_{ud}[s])^{-\kappa}}, \sqrt{(d_{ur}[s])^{-\alpha}}]^{H}\right),\nonumber\\
&R_{ue}[s]=\log_{2}\left(1+[\sqrt{(d_{ue}[s])^{-\kappa}},\sqrt{(d_{ur}[s])^{-\alpha}}]\boldsymbol{\Psi}_{2}\boldsymbol{\Psi}_{2}^{H}\right.\nonumber\\
&\left.[\sqrt{(d_{ue}[s])^{-\kappa}},\sqrt{(d_{ur}[s])^{-\alpha}}]^{H}\right),\nonumber\\
&R_{ut}[s]=\log_{2}\left(1+[\sqrt{(d_{ut}[s])^{-\kappa}},\sqrt{(d_{ur}[s])^{-\alpha}}]\boldsymbol{\Psi}_{3}\boldsymbol{\Psi}_{3}^{H}\right.\nonumber\\
&\left.[\sqrt{(d_{ut}[s])^{-\kappa}},\sqrt{(d_{ur}[s])^{-\alpha}}]^{H}\right).\label{pro59}
\end{align}
\end{small}%
The problem in (\ref{pro55}) is rewritten as
\begin{small}
\begin{subequations}
\begin{align}
\max_{\bar{\mathbf{p}}_{u}[s]}&~\frac{1}{S}\sum_{s=1}^{S}\omega\left(\log_{2}\left(1+[\sqrt{(d_{ud}[s])^{-\kappa}},\sqrt{(d_{ur}[s])^{-\alpha}}]\boldsymbol{\Psi}_{1}\boldsymbol{\Psi}_{1}^{H}\right.\right.\nonumber\\
&\left.\left.[\sqrt{(d_{ud}[s])^{-\kappa}},\sqrt{(d_{ur}[s])^{-\alpha}}]^{H}\right)\right)-\log_{2}\left(1+[\sqrt{(d_{ud}[s])^{-\kappa}},\right.\nonumber\\
&\left.\left.\sqrt{(d_{ur}[s])^{-\alpha}}]\boldsymbol{\Psi}_{2}\boldsymbol{\Psi}_{2}^{H} [\sqrt{(d_{ud}[s])^{-\kappa}},\sqrt{(d_{ur}[s])^{-\alpha}}]^{H}\right)\right)+\nonumber\\
&(1-\omega)\left(\log_{2}\left(1+[\sqrt{(d_{ud}[s])^{-\kappa}},\sqrt{(d_{ur}[s])^{-\alpha}}]\boldsymbol{\Psi}_{1}\boldsymbol{\Psi}_{1}^{H}\right.\right.\nonumber\\
&\left.[\sqrt{(d_{ud}[s])^{-\kappa}},\sqrt{(d_{ur}[s])^{-\alpha}}]^{H}\right)-\log_{2}\left(1+[\sqrt{(d_{ue}[s])^{-\kappa}},\right.\nonumber\\
&\left.\left.\sqrt{(d_{ur}[s])^{-\alpha}}]\boldsymbol{\Psi}_{3}\boldsymbol{\Psi}_{3}^{H}[\sqrt{(d_{ue}[s])^{-\kappa}},\sqrt{(d_{ur}[s])^{-\alpha}}]^{H}\right)\right),\label{pro60a}\\
\mbox{s.t.}~
&(\ref{pro21c}),(\ref{pro21d}),(\ref{pro21g}).&\label{pro60b}
\end{align}\label{pro60}%
\end{subequations}
\end{small}%
We introduce slack variables $\zeta_{1}[s],\zeta_{2}[s],\zeta_{3}[s],\zeta_{4}[s],\zeta_{5}[s],$ $\zeta_{6}[s],\zeta_{7}[s],\zeta_{8}[s],\zeta_{9}[s]$, problem (\ref{pro60}) is rewritten as

\begin{small}
\begin{subequations}
\begin{align}
\max_{{{{\bar{\mathbf{p}}_{u}[s],\zeta_{1}[s],
\atop
\zeta_{2}[s],\zeta_{3}[s],}
\atop
\zeta_{4}[s],\zeta_{5}[s],}
\atop\zeta_{6}[s],\zeta_{7}[s],}\atop\zeta_{8}[s],\zeta_{9}[s]}&~\frac{1}{S}\sum_{s=1}^{S}\omega(\log_{2}(1+\zeta_{1}[s])-\log_{2}(1+\zeta_{2}[s]))\nonumber\\
&+(1-\omega)(\log_{2}(1+\zeta_{1}[s])-\log_{2}(1+\zeta_{3}[s])),\label{pro61a}\\
\mbox{s.t.}~
&[\sqrt{(d_{ud}[s])^{-\kappa}},\sqrt{(d_{ur}[s])^{-\alpha}}]\boldsymbol{\Psi}_{1}\boldsymbol{\Psi}_{1}^{H}\nonumber\\
&[\sqrt{(d_{ud}[s])^{-\kappa}},\sqrt{(d_{ur}[s])^{-\alpha}}]^{H}\geq \zeta_{1}[s],&\label{pro61b}\\
&[\sqrt{(d_{ue}[s])^{-\kappa}},\sqrt{(d_{ur}[s])^{-\alpha}}]\boldsymbol{\Psi}_{2}\boldsymbol{\Psi}_{2}^{H}\nonumber\\
&[\sqrt{(d_{ue}[s])^{-\kappa}},\sqrt{(d_{ur}[s])^{-\alpha}}]^{H}\leq\zeta_{2}[s],&\label{pro61c}\\
&[\sqrt{(d_{ut}[s])^{-\kappa}},\sqrt{(d_{ur}[s])^{-\alpha}}]\boldsymbol{\Psi}_{3}\boldsymbol{\Psi}_{3}^{H}\nonumber\\
&[\sqrt{(d_{ut}[s])^{-\kappa}},\sqrt{(d_{ur}[s])^{-\alpha}}]^{H}\leq\zeta_{3}[s],&\label{pro61d}\\
&\sqrt{(d_{ud}[s])^{-\kappa}}\geq\zeta_{4},\sqrt{(d_{ur}[s])^{-\alpha}}\geq\zeta_{5},&\label{pro61f}\\
&\sqrt{(d_{ue}[s])^{-\kappa}}\leq\zeta_{6},\sqrt{(d_{ur}[s])^{-\alpha}}\leq\zeta_{7},&\label{pro61h}\\
&\sqrt{(d_{ut}[s])^{-\kappa}}\leq\zeta_{8},\sqrt{(d_{ur}[s])^{-\alpha}}\leq\zeta_{9},&\label{pro61j}\\
&\sqrt{(d_{ur}[s])^{-\alpha}}\geq\gamma_{t}^{1/4},&\label{pro61k}
\end{align}\label{pro61}%
\end{subequations}
\end{small}%
where $\gamma_{t}=\frac{\gamma\mathbf{u}^{H}\mathbf{u}\sigma^{2}}{d_{rt}^{-2\alpha}\rho^{2}|\mathbf{u}^{H}[s]\mathbf{H}_{ur}[s]^{H}\boldsymbol{\Theta}[s]\mathbf{A}_{rt}\boldsymbol{\Theta}[s]\mathbf{H}_{ur}[s]\mathbf{w}[s]|^{2}}$.
First, we approximate the objective function in (\ref{pro61a}) by deriving the first order Taylor expansion of $\log_{2}(1+\zeta_{2}[s])$ and $\log_{2}(1+\zeta_{3}[s])$ which is denoted as
\begin{small}
\begin{align}
&\log_{2}(1+\zeta_{2}[s])\leq\log_{2}(1+\zeta_{2,0}[s])+\frac{1}{\ln2(1+\zeta_{2,0}[s])}\nonumber\\
&(\zeta_{2}[s]-\zeta_{2,0}[s]),\label{pro_ver_1}\\
&\log_{2}(1+\zeta_{3}[s])\leq\log_{2}(1+\zeta_{3,0}[s])+\frac{1}{\ln2(1+\zeta_{3,0}[s])}\nonumber\\
&(\zeta_{3}[s]-\zeta_{3,0}[s]).\label{pro_ver_2}
\end{align}
\end{small}%
Therefore, the lower bound of (\ref{pro61a}) is expressed as
\begin{small}
\begin{align}
&\frac{1}{S}\sum_{s=1}^{S}\omega(\log_{2}(1+\zeta_{1}[s])-\log_{2}(1+\zeta_{2}[s]))+(1-\omega)(\log_{2}(1+\nonumber\\
&\zeta_{1}[s])-\log_{2}(1+\zeta_{3}[s]))\geq\frac{1}{S}\sum_{s=1}^{S}\omega(\log_{2}(1+\zeta_{1}[s])-\log_{2}(1+\nonumber\\
&\zeta_{2,0}[s])-\frac{1}{\ln2(1+\zeta_{2,0}[s])}(\zeta_{2}[s]-\zeta_{2,0}[s]))+(1-\omega)(\log_{2}(1+\zeta_{1}[s])\nonumber\\
&-\log_{2}(1+\zeta_{3,0}[s])-\frac{1}{\ln2(1+\zeta_{3,0}[s])}(\zeta_{3}[s]-\zeta_{3,0}[s])).\label{pro_ver_3}
\end{align}
\end{small}%
(\ref{pro_ver_3}) is convex, and (\ref{pro_ver_3}) is used to replace (\ref{pro61a}). Next, we address the non-convexity of the constraints by deriving the first order Taylor expansions of (\ref{pro61b})-(\ref{pro61d}) which are expressed as
\begin{small}
\begin{align}
&-[\zeta_{4,0}[s],\zeta_{5,0}[s]]\boldsymbol{\Psi}_{1}\boldsymbol{\Psi}_{1}^{H} [\zeta_{4,0}[s],\zeta_{5,0}[s]]^{H}\nonumber\\
&+2\mathrm{Re}\left([\zeta_{4,0}[s],\zeta_{5,0}[s]]\boldsymbol{\Psi}_{1}\boldsymbol{\Psi}_{1}^{H}[\zeta_{4}[s],\zeta_{5}[s]]^{H}\right)\geq \zeta_{1}[s],\label{pro_ver_4}\\
&-[\zeta_{6,0}[s],\zeta_{7,0}[s]]\boldsymbol{\Psi}_{2}\boldsymbol{\Psi}_{2}^{H}[\zeta_{6,0}[s],\zeta_{7,0}[s]]^{H}\nonumber\\
&+2\mathrm{Re}\left([\zeta_{6,0}[s],\zeta_{7,0}[s]]\boldsymbol{\Psi}_{2}\boldsymbol{\Psi}_{2}^{H}[\zeta_{6}[s],\zeta_{7}[s]]^{H}\right)\leq\zeta_{2}[s],\label{pro_ver_5}\\
&-[\zeta_{8,0}[s],\zeta_{9,0}[s]]\boldsymbol{\Psi}_{3}\boldsymbol{\Psi}_{3}^{H}[\zeta_{8,0}[s],\zeta_{9,0}[s]]^{H}\nonumber\\
&+2\mathrm{Re}\left([\zeta_{8,0}[s],\zeta_{9,0}[s]]\boldsymbol{\Psi}_{3}\boldsymbol{\Psi}_{3}^{H}[\zeta_{8}[s],\zeta_{9}[s]]^{H}\right)\leq\zeta_{3}[s].\label{pro_ver_6}
\end{align}
\end{small}%
Then, $\sqrt{(d_{ud}[s])^{-\kappa}}\geq \zeta_{4}$, $\sqrt{(d_{ur}[s])^{-\alpha}}\geq \zeta_{5}$, $\sqrt{(d_{ur}[s])^{-\alpha}}\geq \gamma_{t}^{1/4}$, $\sqrt{(d_{ue}[s])^{-\kappa}}\leq \zeta_{6}$, $\sqrt{(d_{ur}[s])^{-\alpha}}\leq \zeta_{7}$, $\sqrt{(d_{ut}[s])^{-\kappa}}\leq \zeta_{8}$, $\sqrt{(d_{ur}[s])^{-\alpha}}\leq \zeta_{9}$ are written as
\begin{small}
\begin{align}
&\left((x_{u}[s]-x_{d}[s])^{2}+(y_{u}[s]-y_{d}[s])^{2}+z_{u}^{2}\right)^{-\kappa/4}\geq\zeta_{4}[s],\label{pro_ver_7}\\
&\left((x_{u}[s]-x_{r})^{2}+(y_{u}[s]-y_{r})^{2}+(z_{u}-z_{r})^{2}\right)^{-\alpha/4}\geq\zeta_{5},\label{pro_ver_8}\\
&\left((x_{u}[s]-x_{r})^{2}+(y_{u}[s]-y_{r})^{2}+(z_{u}-z_{r})^{2}\right)^{-\alpha/4}\geq\gamma_{t}^{1/4},\label{pro_ver_9}\\
&\left((x_{u}[s]-x_{e}[s])^{2}+(y_{u}[s]-y_{e}[s])^{2}+z_{u}^{2}\right)^{-\kappa/4}\leq\zeta_{6}[s],\label{pro_ver_10}\\
&\left((x_{u}[s]-x_{r})^{2}+(y_{u}[s]-y_{r})^{2}+(z_{u}-z_{r})^{2}\right)^{-\alpha/4}\leq\zeta_{7}[s],\label{pro_ver_11}\\
&\left((x_{u}[s]-x_{t}[s])^{2}+(y_{u}[s]-y_{t}[s])^{2}+z_{u}^{2}\right)^{-\kappa/4}\leq\zeta_{8}[s],\label{pro_ver_12}\\
&\left((x_{u}[s]-x_{r})^{2}+(y_{u}[s]-y_{r})^{2}+(z_{u}-z_{r})^{2}\right)^{-\alpha/4}\leq\zeta_{9}[s].\label{pro_ver_13}
\end{align}
\end{small}%
The expressions in (\ref{pro_ver_7})-(\ref{pro_ver_13}) can be further approximated as
\begin{small}
\begin{align}
&x_{u}^{2}[s]+x_{d}^{2}[s]-2x_{u}[s]x_{d}[s]+y_{u}^{2}[s]+y_{d}^{2}[s]-2y_{u}[s]y_{d}[s]+z_{u}^{2}\nonumber\\
&-\zeta_{4}[s]^{-\frac{4}{\kappa}}\leq 0,\label{pro_ver_14}\\
&x_{u}^{2}[s]+x_{r}^{2}-2x_{u}[s]x_{r}+y_{u}^{2}[s]+y_{r}^{2}-2y_{u}[s]y_{r}+z_{u}^{2}+z_{r}^{2}-2z_{u}z_{r}\nonumber\\
&-\zeta_{5}[s]^{-\frac{4}{\alpha}}\leq 0,\label{pro_ver_15}\\
&x_{u}^{2}[s]+x_{r}^{2}-2x_{u}[s]x_{r}+y_{u}^{2}[s]+y_{r}^{2}-2y_{u}[s]y_{r}+z_{u}^{2}+z_{r}^{2}-2z_{u}z_{r}\nonumber\\
&-\gamma_{t}^{-\frac{1}{16\alpha}}\leq 0,\label{pro_ver_16}\\
&\zeta_{6}[s]^{-\frac{4}{\kappa}}-x_{u}^{2}[s]-x_{e}^{2}[s]+2x_{u}[s]x_{e}[s]-y_{u}^{2}[s]-y_{e}^{2}[s]+2y_{u}[s]y_{e}[s]\nonumber\\
&-z_{u}^{2}\geq 0,\label{pro_ver_17}\\
&\zeta_{7}[s]^{-\frac{4}{\alpha}}-x_{u}^{2}[s]-x_{r}^{2}+2x_{u}[s]x_{r}-y_{u}^{2}[s]-y_{r}^{2}+2y_{u}[s]y_{r}-z_{u}^{2}-z_{r}^{2}\nonumber\\
&+2z_{u}z_{r}\geq 0,\label{pro_ver_18}\\
&\zeta_{8}[s]^{-\frac{4}{\kappa}}-x_{u}^{2}[s]-x_{t}^{2}[s]+2x_{u}[s]x_{t}[s]-y_{u}^{2}[s]-y_{t}^{2}[s]+2y_{u}[s]y_{t}[s]\nonumber\\
&-z_{u}^{2}\geq 0,\label{pro_ver_19}\\
&\zeta_{9}[s]^{-\frac{4}{\alpha}}-x_{u}^{2}[s]-x_{r}^{2}+2x_{u}[s]x_{r}-y_{u}^{2}[s]-y_{r}^{2}+2y_{u}[s]y_{r}-z_{u}^{2}-z_{r}^{2}\nonumber\\
&+2z_{u}z_{r}\geq 0,\label{pro_ver_20}
\end{align}
\end{small}%
in which (\ref{pro_ver_16}) is convex. Since (\ref{pro_ver_14}) is a non-convex constraint, we use the first order Taylor expansion of $\zeta_{4}[s]^{-\frac{4}{\kappa}}$ to approximate (\ref{pro_ver_14}) which is given by 
\begin{small}
\begin{align}
\zeta_{4}[s]^{-\frac{4}{\kappa}}\geq \zeta_{4,0}[s]^{-\frac{4}{\kappa}}-\frac{4}{\kappa}\zeta_{4,0}[s]^{-\frac{4}{\kappa}-1}(\zeta_{4}[s]-\zeta_{4,0}[s]). \label{pro_ver_21}
\end{align}
\end{small}%
Based on (\ref{pro_ver_21}), constraint (\ref{pro_ver_14}) is rewritten as
\begin{small}
\begin{align}
&x_{u}^{2}[s]+x_{d}^{2}[s]-2x_{u}[s]x_{d}[s]+y_{u}^{2}[s]+y_{d}^{2}[s]-2y_{u}[s]y_{d}[s]+z_{u}^{2}\nonumber\\
&-\zeta_{4,0}[s]^{-\frac{4}{\kappa}}+\frac{4}{\kappa}\zeta_{4,0}[s]^{-\frac{4}{\kappa}-1}(\zeta_{4}[s]-\zeta_{4,0}[s])\leq 0, \label{pro_ver_22} 
\end{align}
\end{small}%
which is convex. Similarly, constraint (\ref{pro_ver_15}) is rewritten as
\begin{small}
\begin{align}
&x_{u}^{2}[s]+x_{r}^{2}-2x_{u}[s]x_{r}+y_{u}^{2}[s]+y_{r}^{2}-2y_{u}[s]y_{r}+z_{u}^{2}+z_{r}^{2}\nonumber\\
&-2z_{u}z_{r}-\zeta_{5,0}[s]^{-\frac{5}{\kappa}}+\frac{5}{\kappa}\zeta_{5,0}[s]^{-\frac{5}{\kappa}-1}(\zeta_{5}[s]-\zeta_{5,0}[s])\leq 0, \label{pro_ver_23}  
\end{align}
\end{small}%
which is also convex. The first order Taylor expansions of $-x_{u}^{2}[s]$ and $-y_{u}^{2}[s]$ are given by 
\begin{small}
\begin{align}
&-x_{u}[s]^{2}\leq x_{u,0}^{2}[s]-2x_{u,0}[s]x_{u}[s], -y_{u}[s]^{2}\leq y_{u,0}^{2}[s]\nonumber\\
&-2y_{u,0}[s]y_{u}[s][s].\label{pro_ver_24}  
\end{align}
\end{small}%
Constraints (\ref{pro_ver_17})-(\ref{pro_ver_20}) are rewritten as
\begin{small}
\begin{align}
&\zeta_{6}[s]^{-\frac{4}{\kappa}}+x_{u,0}^{2}[s]-2x_{u,0}[s]x_{u}[s]-x_{e}^{2}[s]+2x_{u}[s]x_{e}[s]+y_{u,0}^{2}[s]\nonumber\\
&-2y_{u,0}[s]y_{u}[s]-y_{e}^{2}[s]+2y_{u}[s]y_{e}[s]-z_{u}^{2}\geq 0,\label{pro_ver_25}\\
&\zeta_{7}[s]^{-\frac{4}{\alpha}}+x_{u,0}^{2}[s]-2x_{u,0}[s]x_{u}[s]-x_{r}^{2}+2x_{u}[s]x_{r}+y_{u,0}^{2}[s]\nonumber\\
&-2y_{u,0}[s]y_{u}[s]-y_{r}^{2}+2y_{u}[s]y_{r}-z_{u}^{2}-z_{r}^{2}+2z_{u}z_{r}\geq 0,\label{pro_ver_26}\\
&\zeta_{8}[s]^{-\frac{4}{\kappa}}+x_{u,0}^{2}[s]-2x_{u,0}[s]x_{u}[s]-x_{t}^{2}[s]+2x_{u}[s]x_{t}[s]\nonumber\\
&+y_{u,0}^{2}[s]-2y_{u,0}[s]y_{u}[s]-y_{t}^{2}[s]+2y_{u}[s]y_{t}[s]-z_{u}^{2}\geq 0,\label{pro_ver_27}\\
&\zeta_{9}[s]^{-\frac{4}{\alpha}}+x_{u,0}^{2}[s]-2x_{u,0}[s]x_{u}[s]-x_{r}^{2}+2x_{u}[s]x_{r}+y_{u,0}^{2}[s]\nonumber\\
&-2y_{u,0}[s]y_{u}[s]-y_{r}^{2}+2y_{u}[s]y_{r}-z_{u}^{2}-z_{r}^{2}+2z_{u}z_{r}\geq 0,\label{pro_ver_28}
\end{align}
\end{small}%
which are convex. Finally, the ISAC-UAV trajectory optimization in (\ref{pro61}) can be approximated as
\begin{small}
\begin{subequations}
\begin{align}
\max_{{{\bar{\mathbf{p}}_{u}[s],\zeta_{1}[s],\zeta_{2}[s],
\atop
\zeta_{3}[s],\zeta_{4}[s],\zeta_{5}[s],}
\atop
\zeta_{6}[s],\zeta_{7}[s],\zeta_{8}[s],}
\atop\zeta_{9}[s]}&~(\ref{pro_ver_3}),\label{pro63a}\\
\mbox{s.t.}~
&(\ref{pro_ver_4}),(\ref{pro_ver_5}),(\ref{pro_ver_6}),(\ref{pro_ver_22}),(\ref{pro_ver_23}),(\ref{pro_ver_25}),(\ref{pro_ver_26}),(\ref{pro_ver_27}),(\ref{pro_ver_28}),&\label{pro63b}
\end{align}\label{pro63}%
\end{subequations}
\end{small}%
which is convex.

\subsection{ISAC-UAV Receive Beamforming Optimization}
Finally, we consider optimizing the ISAC receive beamforming for a fixed UAV trajectory, RIS phase shift matrix and transmit beamforming vector. We have
\begin{small}
\begin{subequations}
\begin{align}
\max_{\mathbf{u}[s]}&~\frac{1}{S}\sum_{s=1}^{S}R_{sum}[s],\label{pro64a}\\
\mbox{s.t.}~
&(\ref{pro21g}),(\ref{pro21h}),&\label{pro64b}
\end{align}\label{pro64}%
\end{subequations}
\end{small}%
Based on \cite{9390351}, receive beamforming $\mathbf{u}[s]$ is not involved in $\mathrm{SNR}_{ud}[s]$ and $\mathrm{SNR}_{ue}[s]$, and the optimization objective is only relevant to $\mathrm{SNR}_{ut}[s]$. Thus, problem (\ref{pro64}) is rewritten as
\begin{small}
\begin{subequations}
\begin{align}
\min_{\mathbf{u}[s]}&~\frac{1}{S}\sum_{s=1}^{S}\mathbf{u}^{H}[s]\boldsymbol{\Omega}[s]\mathbf{u}[s],\label{pro65a}\\
\mbox{s.t.}~
&\mathbf{u}^{H}[s]\boldsymbol{\Omega}\mathbf{u}[s]\geq\gamma\sigma^{2},(\ref{pro21h}),&\label{pro65b}
\end{align}\label{pro65}%
\end{subequations}
\end{small}%
where $\boldsymbol{\Omega}[s]=L_{urt}^{4}[s]\mathbf{H}_{ur}[s]^{H}\boldsymbol{\Theta}[s]\mathbf{A}_{rt}\boldsymbol{\Theta}[s]\mathbf{H}_{ur}[s]\mathbf{w}[s]\mathbf{w}[s]^{H}$ $\mathbf{H}_{ur}[s]^{H}\boldsymbol{\Theta}[s]^{H}\mathbf{A}_{rt}^{H}\boldsymbol{\Theta}[s]^{H}\mathbf{H}_{ur}[s]$.
According to \cite{7093191}, we use the Majorize-Minimization(MM) algorithm to resolve problem (\ref{pro65}) and problem (\ref{pro65}) is rewritten as
\begin{small}
\begin{subequations}
\begin{align}
\min_{\mathbf{u}[s]}&~\frac{1}{S}\sum_{s=1}^{S}\mathbf{u}^{H}[s]\lambda_{max}(\boldsymbol{\Omega})\mathbf{u}[s]+2\mathrm{Re}(\mathbf{u}^{H}[s](\boldsymbol{\Omega}-\lambda_{max}(\boldsymbol{\Omega}))\mathbf{I})\nonumber\\
&\mathbf{u}_{0}^{H}[s],\label{pro66a}\\
\mbox{s.t.}~
&\mathbf{u}^{H}[s]\lambda_{max}(\boldsymbol{\Omega})\mathbf{u}[s]+2\mathrm{Re}(\mathbf{u}^{H}[s](\boldsymbol{\Omega}-\lambda_{max}(\boldsymbol{\Omega}))\mathbf{I})\mathbf{u}_{0}^{H}[s]\nonumber\\
&\geq\gamma\sigma^{2},(\ref{pro21h}),&\label{pro66b}
\end{align}\label{pro66}%
\end{subequations}
\end{small}%
Since $\|\mathbf{u}[s]\|_{2}=1$, problem (\ref{pro66}) is reformulated as
\begin{small}
\begin{subequations}
\begin{align}
\min_{\mathbf{u}[s]}&~\frac{1}{S}\sum_{s=1}^{S}2\mathrm{Re}(\mathbf{u}^{H}[s](\boldsymbol{\Omega}-\lambda_{max}(\boldsymbol{\Omega}))\mathbf{I})\mathbf{u}_{0}^{H}[s],\label{pro67a}\\
\mbox{s.t.}~
&\lambda_{max}(\boldsymbol{\Omega}+2\mathrm{Re}(\mathbf{u}^{H}[s](\boldsymbol{\Omega}-\lambda_{max}(\boldsymbol{\Omega}))\mathbf{I})\mathbf{u}_{0}^{H}[s]\geq\gamma\sigma^{2},(\ref{pro21h}),&\label{pro67b}
\end{align}\label{pro67}%
\end{subequations}
\end{small}%
The constraint condition in (\ref{pro21h}) is equivalently transformed as $\|\mathbf{u}[s]\|_{2}\leq 1$, $\|\mathbf{u}[s]\|_{2}\geq 1$, thus, we have
\begin{small}
\begin{subequations}
\begin{align}
\min_{\mathbf{u}[s]}&~\frac{1}{S}\sum_{s=1}^{S}2\mathrm{Re}(\mathbf{u}^{H}[s](\boldsymbol{\Omega}-\lambda_{max}(\boldsymbol{\Omega}))\mathbf{I})\mathbf{u}_{0}^{H}[s],\label{pro68a}\\
\mbox{s.t.}~
&\lambda_{max}(\boldsymbol{\Omega}+2\mathrm{Re}(\mathbf{u}^{H}[s](\boldsymbol{\Omega}-\lambda_{max}(\boldsymbol{\Omega}))\mathbf{I})\mathbf{u}_{0}^{H}[s]\geq\gamma\sigma^{2},&\nonumber\\
&\|\mathbf{u}\|_{2}\leq 1, \|\mathbf{u}\|_{2}\geq 1,&\label{pro68b}
\end{align}\label{pro68}%
\end{subequations}
\end{small}%
$\|\mathbf{u}\|_{2}\geq 1$ can be approximated as $\|\mathbf{u}_{0}\|_{2}+2\mathrm{Re}(\mathbf{u}_{0}^{H}(\mathbf{u}-\mathbf{u}_{0}))\geq 1$. Problem (\ref{pro68}) is rewritten as
\begin{small}
\begin{subequations}
\begin{align}
\min_{\mathbf{u}[s]}&~\frac{1}{S}\sum_{s=1}^{S}2\mathrm{Re}(\mathbf{u}^{H}[s](\boldsymbol{\Omega}-\lambda_{max}(\boldsymbol{\Omega}))\mathbf{I})\mathbf{u}_{0}^{H}[s],\label{pro69a}\\
\mbox{s.t.}~
&\lambda_{max}(\boldsymbol{\Omega}+2\mathrm{Re}(\mathbf{u}^{H}[s](\boldsymbol{\Omega}-\lambda_{max}(\boldsymbol{\Omega}))\mathbf{I})\mathbf{u}_{0}^{H}[s]\geq\gamma\sigma^{2},&\label{pro69b}\\
&\|\mathbf{u}\|_{2}\leq 1,&\label{pro69c}\\
&\|\mathbf{u}_{0}\|_{2}+2\mathrm{Re}(\mathbf{u}_{0}^{H}(\mathbf{u}-\mathbf{u}_{0}))\geq 1,&\label{pro69d}
\end{align}\label{pro69}%
\end{subequations}
\end{small}%
which is convex and can be solved using CVX.
Finally, the proposed BCD algorithm is summarized in \textbf{Algorithm~\ref{algo1}}
\begin{algorithm}%
\caption{Proposed BCD Algorithm for Problem (\ref{pro22})} \label{algo1}
\hspace*{0.02in}{\bf Initialize:}
$\bar{\mathbf{p}}_{u}^{(0)}[s]$, $\boldsymbol{\Theta}^{(0)}[s]$, $\mathbf{w}^{(0)}[s]$, $\mathbf{u}^{(0)}[s]$.\\
\hspace*{0.02in}{\bf Repeat:}~$t=t+1$.\\
Given $\bar{\mathbf{p}}_{u}^{(t)}[s]$, $\boldsymbol{\Theta}^{(t)}[s]$, $\mathbf{w}^{(t)}[s]$, $\mathbf{u}^{(t)}[s]$, computing $\Delta\mathbf{h}_{e1}^{t}[s]$ and $\Delta\mathbf{h}_{t1}^{t}[s]$ based on (\ref{pro37});\\
Given $\bar{\mathbf{p}}_{u}^{(t)}[s]$, $\boldsymbol{\Theta}^{(t)}[s]$, $\mathbf{u}^{(t)}[s]$, find optimal solution $\mathbf{w}^{*}[s]$ of problem (\ref{pro43}). Update $\mathbf{w}^{(t+1)}[s]=\mathbf{w}^{*}[s]$;\\
Given $\bar{\mathbf{p}}_{u}^{(t)}[s]$, $\mathbf{w}^{(t+1)}[s]$, $\mathbf{u}^{(t)}[s]$, find optimal solution $\boldsymbol{\Theta}^{*}[s]$ of problem (\ref{pro54}). Update $\boldsymbol{\Theta}^{(t+1)}[s]=\boldsymbol{\Theta}^{*}[s]$;\\
Given $\boldsymbol{\Theta}^{(t+1)}[s]$, $\mathbf{w}^{(t+1)}[s]$, $\mathbf{u}^{(t)}[s]$, find optimal solution $\boldsymbol{\Theta}^{*}[s]$ of problem (\ref{pro63}). Update $\bar{\mathbf{p}}_{u}^{(t+1)}[s]=\bar{\mathbf{p}}_{u}^{(*)}[s]$;\\
Given $\boldsymbol{\Theta}^{(t+1)}[s]$, $\mathbf{w}^{(t+1)}[s]$, $\bar{\mathbf{p}}_{u}^{(t+1)}[s]$, find optimal solution $\mathbf{u}^{*}[s]$ of problem (\ref{pro69}). Update $\mathbf{u}^{(t+1)}[s]=\mathbf{u}^{*}[s]$;\\
\hspace*{0.02in}{\bf Until:}~$|R_{sum}^{(t+1)}[s]-R_{sum}^{(t)}[s]|\leq \nu$.\\
\hspace*{0.02in}{\bf Output:}
$\boldsymbol{\Theta}^{(t+1)}[s]$, $\mathbf{w}^{(t+1)}[s]$, $\bar{\mathbf{p}}_{u}^{(t+1)}[s]$,$\mathbf{u}^{(t+1)}[s]$.\\
\end{algorithm}

\section{Comuputational Complexity and Convergence Analysis}\label{IV}
In this section, we discuss the computation complexity of the proposed algorithm. The solution problem (\ref{pro37}) is obtained by using the times operations, so the computational complexity of problem (\ref{pro37}) is $\mathcal{O}(2M+1)$. Since the SCA algorithm and cvx is used for problem (\ref{pro45}) by using interior point method, the computational complexity of problem (\ref{pro45}) is denoted as$\mathcal{O}((N_{t}+1)^{3.5})$. Similarly, the computational complexity of using the interior point method for solving problem (\ref{pro54}), problem (\ref{pro63}) and problem (\ref{pro69}), therefore, their computational complexity are  $\mathcal{O}((3)^{3.5})$ and $\mathcal{O}((N_{r}+1)^{3.5})$ and $\mathcal{O}((N_{T}+1)^{3.5})$, respectively. Finally, we assume that the number of iterations for the proposed algorithm is $\bar{t}$, so the computational complexity of the proposed algorithm can be expressed as
$\mathcal{O}(\bar{t}((2M+1)+(N_{r}+1)^{3.5}+(3)^{3.5}+(N_{r}+1)^{3.5}+(N_{T}+1)^{3.5}))$.
Next, we discuss the convergence of the proposed algorithm, and the proof is given in \cite{10168298}.


\section{Numerical Results}\label{V}
\begin{figure}[htbp]
\centering
\begin{minipage}[t]{0.48\textwidth}
\centering
\includegraphics[scale=0.5]{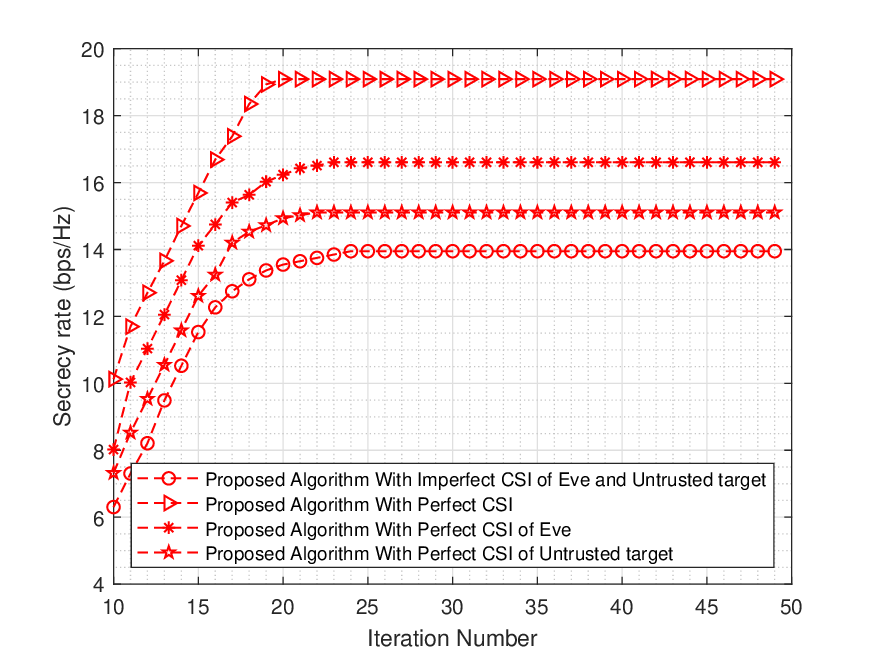}
\caption{Average secrecy rate versus the number of iterations.}
\label{FIGURE1}
\end{minipage}
\begin{minipage}[t]{0.48\textwidth}
\centering
\includegraphics[scale=0.5]{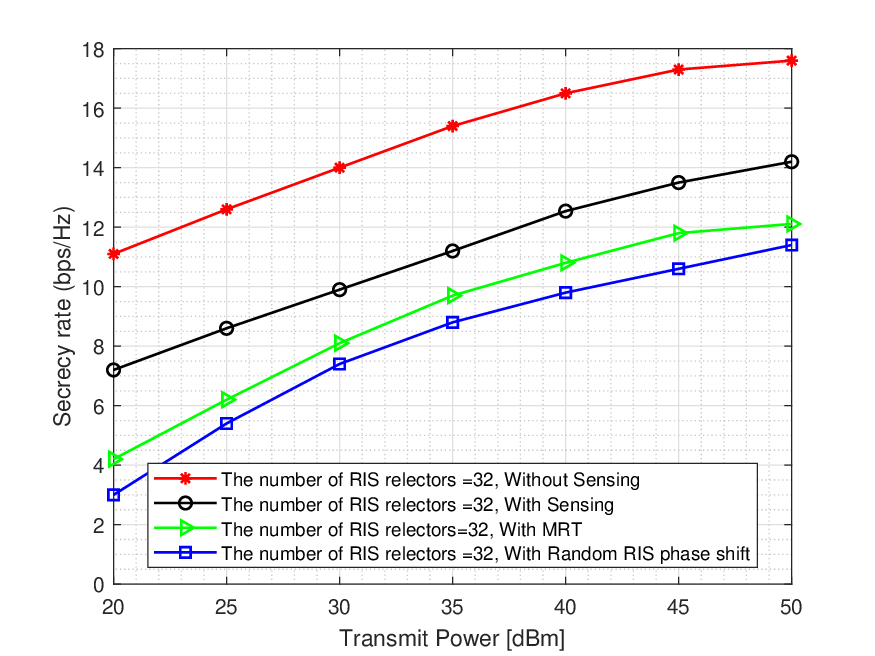}
\caption{Average secrecy rate versus transmit power with different algorithms.}
\label{FIGURE2}
\end{minipage}
\end{figure}
In this section, we present numerical examples to highlight the secrecy performance of the RIS-aided ISAC-UAV communication system. 
The positions of RIS, IoT device, Eve and UT is set as $(0,20,10)$, $(10,20,0)$, $(30,25,0)$, $(50,30,0)$ in meters. The initial UAV position is $(0,0,15)$ and the final UAV position is $(60,30,10)$ in meters. The maximum speed of the UAV is $60m/s$ with slot length $\Delta_{s}=0.8s$. The path loss coefficient and the
path loss exponents $\kappa=3.1$, $\alpha=2.5$ The Rician factor of UAV-to-IoT device is $\beta_{ud}=15$dB due to a strong line-of-sight path. The Rician factor of UAV-to-RIS, UAV-to-Eve, RIS-to-Untrusted target, RIS-to-Eve, RIS-to-IoT device are $\beta_{ur}=5$dB, $\beta_{ue}=5$dB, $\beta_{ut}=5$dB, $\beta_{re}=5$dB, $\beta_{rt}=5$dB. The max transmit power $P=30$dBm and the peak power $P_{peak}=90$dBm.
Fig.~\ref{FIGURE1} shows the convergence behavior of the proposed algorithm. 
Fig.~\ref{FIGURE1} shows the convergence curve of robust channel (communication secrecy rate during iteration) when $N_{t}= 64$, $N_{R}= 16$, $\gamma= 0$ dB. 
The results show that all the curves increase monotonously and converge to a certain upper bound from the initial point, which proves the convergence of the proposed algorithm. As expected, we see that the secrecy rate increases with more available CSI for Eve and UT. Furthermore, the convergence rate of the algorithm is similar for both perfect and imperfect CSI which highlights the robustness of our algorithm.

Fig.~\ref{FIGURE2} depicts the relationship between transmit power and the secrecy rate of the RIS-assisted ISAC-UAV communication system. The simulation results show that as the transmission power increases, the system secrecy rate increases. This is because after the transmission power is increased, more energy will be used to improve the information transmission rate of IoT devices and suppress Eve. Furthermore, in this section, a comparison of several methods is considered. The first takes the lack of sensing rate constraint as an upper bound. In the second case, Eve is not considered, and the maximum ratio transmission (MRT) method is directly used to process the transmit beamforming. The optimization of other variables uses the algorithm mentioned in this paper. In the third case, the RIS phase shift is randomly selected between $0$ and $2\pi$, and other variables are optimized using the algorithm mentioned in this paper. It is easy to find from Fig.~\ref{FIGURE2} that the performance of the proposed algorithm is closer to the upper bound. The RIS phase shift method based on stochastic optimization and the transmit beamforming design based on MRT have poor performance. This is because stochastic optimization cannot fully utilize the beamforming gain of RIS, whereas the MRT method does not consider the suppression of Eve in the user transmission. 

\begin{figure}[htbp]
\centering
\begin{minipage}[t]{0.48\textwidth}
\centering
\includegraphics[scale=0.5]{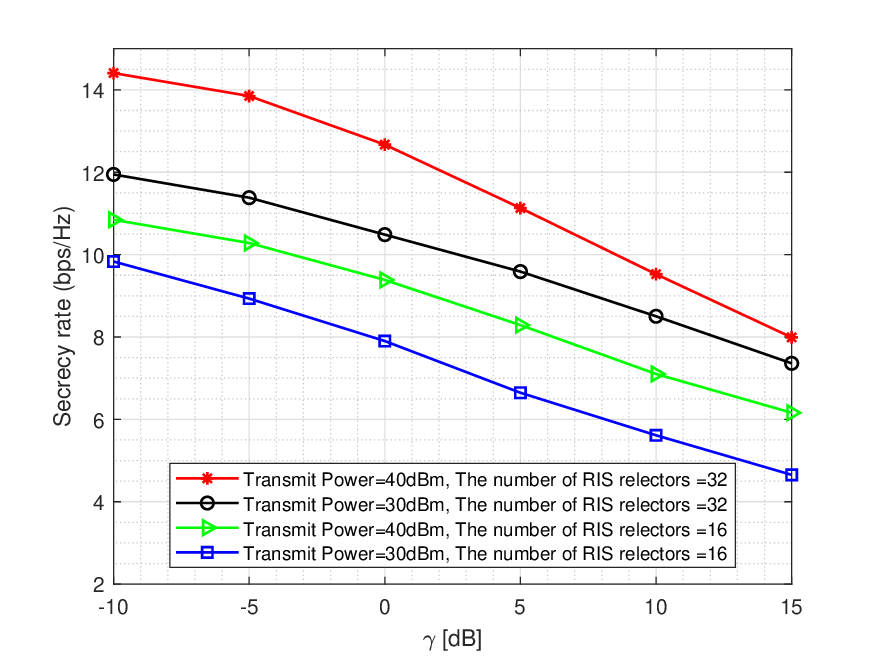}
\caption{Average secrecy rate versus $\gamma$ with different transmit power and the number of RIS reflectors.}
\label{FIGURE3}
\end{minipage}
\begin{minipage}[t]{0.48\textwidth}
\centering
\includegraphics[scale=0.5]{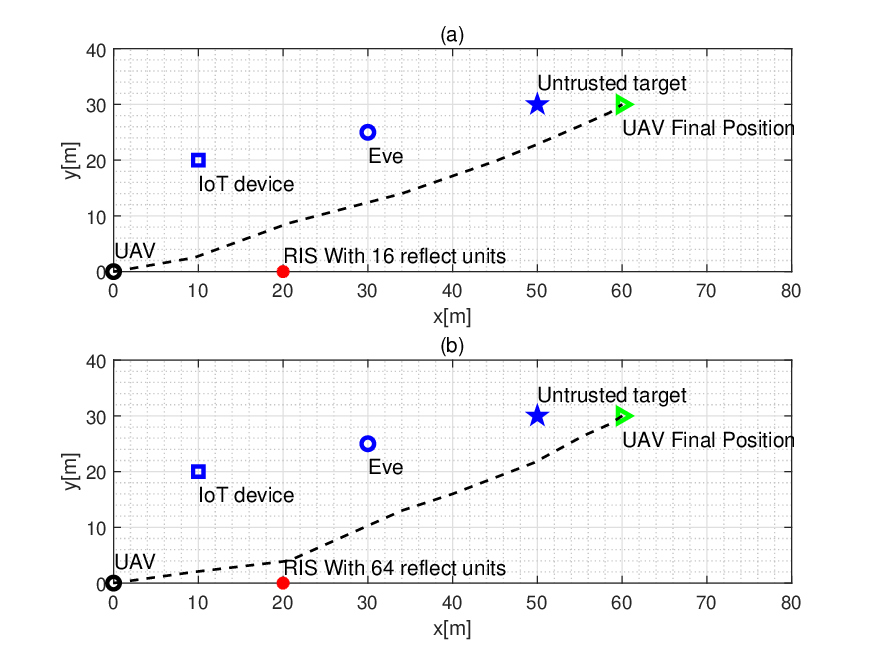}
\caption{UAV trajectory versus $N_{R}$ optimization with different sensing rate constraints where $N_{R}=16$ in (a) where average secrecy rate
is $17.1$~bps/Hz and $N_{R}=64$ in (b) where average secrecy rate
is $20.3$~bps/Hz.}
\label{FIGURE4}
\end{minipage}
\end{figure}

To show the relationship between communication security rate and sensing performance, as shown in Fig.~\ref{FIGURE3}, we plot the relationship between sensing rate limit A and secure sensing rate performance. It is easy to find from Fig.~\ref{FIGURE3} that as the sensing rate limit A increases, the secrecy rate of the system will show a downward trend, which shows that the improvement of sensing performance must sacrifice the secrecy rate of the system. When other resources are fixed, a high sensing rate limit will occupy more resources to improve sensing accuracy. When the sensing rate increases again, you will find that the system secrecy rate decreases faster. This is because of the existence of RIS; the sensing rate is mainly composed of two parts. One part is the power transmitted back to the UAV after the target reflects the RIS; the other is the UAV power—the power of the transmitted signal reaching the target via RIS. When the sensing rate is at a low limit, both parts can meet the sensing rate limit of the system. However, when the sensing rate limit is further increased, the power of both parts cannot be satisfied. Therefore, more beam directions are required to initially point to the target to maintain the sensing rate, which will cause the communication rate of the IoT device to be reduced, further reducing the system's secure communication rate. In addition, Fig.~\ref{FIGURE3} also depicts the relationship between the number of RIS reflection units and the secrecy rate of RIS-assisted ISAC-UAV communication. It is easy to find from Fig.~\ref{FIGURE3} that as the number of RIS reflection units increases, the system communication secrecy rate will also increase, which shows that RIS is still effective in improving the secrecy rate of the ISAC-UAV communication system.

In Fig.~\ref{FIGURE4}, we assume the trajectory of the UAV running in the scenario where T=100 seconds and the UAV speed is 60m/s. Fig.~\ref{FIGURE4}(a), it can be seen that the UAV first flies to a position close to the IRS and keeps hovering at this position. After that, the UAV flies to the final point in a line. Compared with Fig.~\ref{FIGURE4}(a), it can be seen from Fig.~\ref{FIGURE4}(b) that due to the increase in the number of RIS reflection units, the UAV flies to a position closer to the RIS to obtain higher gain and remains hovering at this position. After that, the UAV moves in a line to fly to the final point. This is because the RIS reflection unit brings greater power gain. Therefore, to achieve a higher power gain, the UAV trajectory will be closer to RIS. In addition, the trajectory of the UAV is not as close as possible to the IoT device, because the proposed algorithm balances the communication rate and suppresses Eve.

\begin{figure}[htbp]
\centering
\begin{minipage}[t]{0.49\textwidth}
\centering
\includegraphics[scale=0.5]{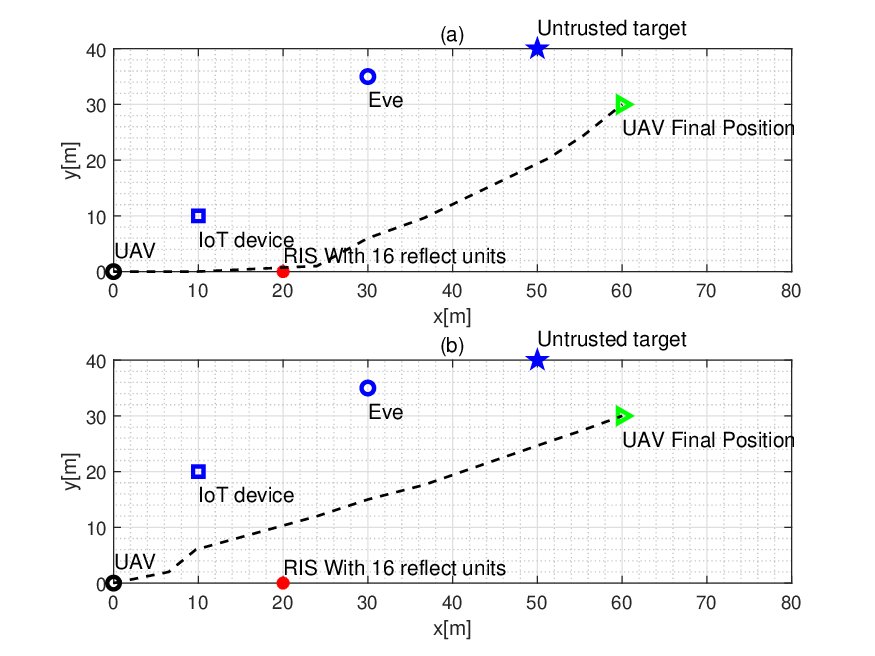}
\caption{UAV trajectory versus position of IoT device optimization with IoT device position (10,10)~m in (a) where average secrecy rate
is $18.7$~bps/Hz and IoT device position (10,20)~m in (b) where average secrecy rate
is $16.1$~bps/Hz.}
\label{FIGURE5}
\end{minipage}
\begin{minipage}[t]{0.49\textwidth}
\centering
\includegraphics[scale=0.5]{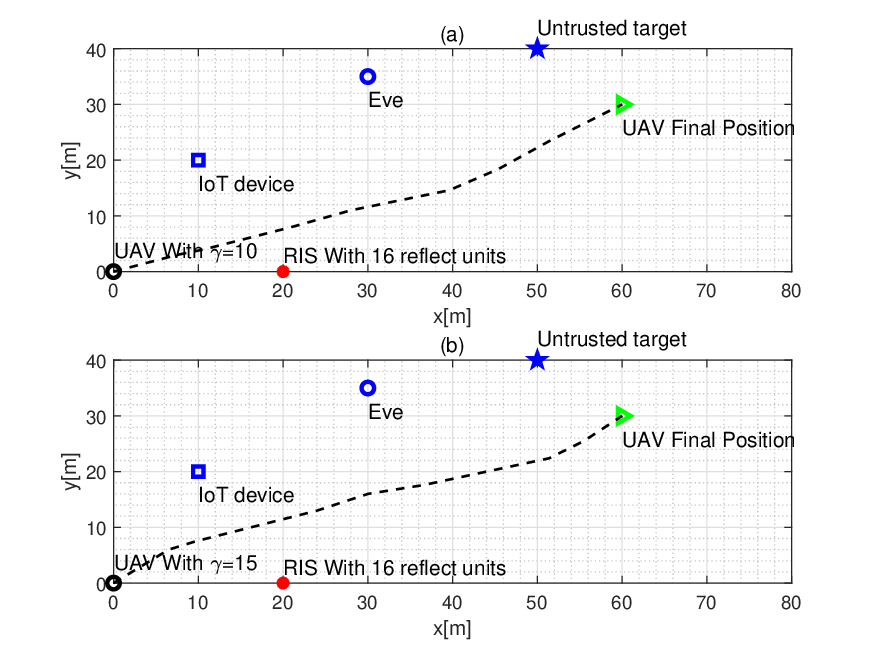}
\caption{UAV trajectory versus $\gamma$ optimization with different sensing rate constraints where $\gamma=10$~dB in (a) where average secrecy rate
is $16.3$~bps/Hz and $\gamma=15$~dB in (b) where average secrecy rate
is $12.8$~bps/Hz.}
\label{FIGURE6}
\end{minipage}
\end{figure}

In Figure~\ref{FIGURE5}, we consider the UAV's trajectory with T=100 seconds and the UAV speed is 5m/s. In Figure~\ref{FIGURE5}(a), the UAV first flies to a position close to the IRS, then keeps hovering, and finally flies directly to the endpoint. Comparing Figure~\ref{FIGURE5}(a) and Figure~\ref{FIGURE5}(b), we see that when the IoT device is far from the RIS, the UAV will be closer to the IoT device. This is because when the IoT device is far away from the RIS, the path loss from the UAV to the IoT device increases, and the power gain of the RIS cannot make up for this loss so the UAV will be closer to the IoT device and farther away from the RIS.

In Figure~\ref{FIGURE6}, we consider that the trajectory of the UAV is with $T=100$ seconds, and the speed of the UAV is $60$m/s. When the sensing rate is limited to 10, as shown in Fig.~\ref{FIGURE6}(a), the UAV first flies to a position close to the UT, then keeps hovering, and finally flies to the end point. Comparing Fig.~\ref{FIGURE6}(a) and Fig.~\ref{FIGURE6}(b), we see that when the sensing rate constraint $\gamma$ is increased to $15$, the UAV will fly even closer to the UT because when the sensing rate constraint $\gamma$ increases, the reflection gain of the RIS and the UAV transmit power are difficult to meet the sensing rate constraint. Therefore, to meet the sensing rate constraint, the UAV's flight trajectory must weigh the communication confidentiality rate and sensing rate, and therefore UAV cannot be completely close to the RIS or IoT device. However, this will cause Eve's gain to increase as well. Therefore, this also shows that increasing the sensing rate constraint $\gamma$ will sacrifice the secrecy rate of the communication system. In actual system design, a trade-off must be made between the security rate and the sensing rate.

\section{Conclusion and Future Works}\label{VI}
In this paper, we propose a novel UAV-ISAC system where RIS is employed to improve the secrecy rate performance for a ground IoT device in the presence of Eve and UT. We consider that when transmitting information and sensing targets between the UAV, IoT device, and UT, it is possible for Eve to eavesdrop on information and UT to obtain legitimate information.  Therefore, we propose a new algorithm to optimize the UAV trajectories, CSI robust design, passive beamforming of RIS, and receiver transmit and receive beamforming to maximize the RIS-aided ISAC-UAV's average secrecy rate. This problem is a non-convex problem due to multi-variable coupling. The BCD algorithm based on SCA and MM is proposed to solve this non-convex problem. Simulation results show that our proposed algorithm converges and that the RIS can help to balance communication security rate and target sensing accuracy.
Furthermore, the robustness of our proposed algorithm is demonstrated against inaccurate estimates of eavesdropping channel CSI. 
\appendices
\section{Proof of \ref{pro53}}\label{appA}
According to\cite{franklin2012matrix}, $\mathrm{tr}(\mathbf{C}_{1}[s]\tilde{\mathbf{Q}}[s]\mathbf{C}_{2}[s]\tilde{\mathbf{Q}}[s])$ can be rewritten as
\begin{small}
\begin{align}
&\mathrm{tr}(\mathbf{C}_{1}[s]\tilde{\mathbf{Q}}[s]\mathbf{C}_{2}[s]\tilde{\mathbf{Q}}[s])=\left(\mathrm{vec}\left(\tilde{\mathbf{Q}}^{T}[s]\right)\right)^{T}(\mathbf{C}_{2}^{T}[s]\otimes\mathbf{C}_{1}[s])\nonumber\\
&\mathrm{vec}(\tilde{\mathbf{Q}}[s]).\label{proA1_1}
\end{align}
\end{small}%
We continue to use vector operation and have
\begin{small}
\begin{align}
&\left(\mathrm{vec}\left(\tilde{\mathbf{Q}}^{T}[s]\right)\right)^{T}(\mathbf{C}_{2}^{T}[s]\otimes\mathbf{C}_{1}[s])\mathrm{vec}(\tilde{\mathbf{Q}}[s])=(\boldsymbol{\theta}^{T}[s]\otimes\boldsymbol{\theta}^{H}[s])\nonumber\\
&(\mathbf{C}_{2}^{T}[s]\otimes\mathbf{C}_{1}[s])(\boldsymbol{\theta}^{*}[s]\otimes\boldsymbol{\theta}[s])=(\boldsymbol{\theta}^{T}[s]\otimes\boldsymbol{\theta}^{H}[s])\hat{\boldsymbol{\Xi}}(\boldsymbol{\theta}^{*}[s]\otimes\boldsymbol{\theta}[s]),\label{proA1_2}
\end{align}
\end{small}%
where $\hat{\boldsymbol{\Xi}}[s]=\mathbf{C}_{2}^{T}[s]\otimes\mathbf{C}_{1}[s]\succeq\mathbf{0}$. Since $\mathbf{C}_{1}[s]$ and $\mathbf{C}_{2}[s]$ are the Hermitian matrix, $\hat{\boldsymbol{\Xi}}[s]=\mathbf{C}_{2}^{*}[s]\otimes\mathbf{C}_{1}^{H}[s]=\mathbf{C}_{2}^{T}[s]\otimes\mathbf{C}_{1}[s]$.
Then, we use eigenvalue decomposition for $\hat{\boldsymbol{\Xi}}[s]$ and $\hat{\boldsymbol{\Xi}}[s]=\mathbf{U}[s]\bar{\boldsymbol{\Xi}}[s]\mathbf{U}^{H}[s]$, thus $(\boldsymbol{\theta}^{T}[s]\otimes\boldsymbol{\theta}^{H}[s])\hat{\boldsymbol{\Xi}}[s](\boldsymbol{\theta}^{*}[s]\otimes\boldsymbol{\theta}[s])=\left\|(\boldsymbol{\theta}^{T}[s]\otimes\boldsymbol{\theta}^{H}[s])\mathbf{U}[s]\bar{\boldsymbol{\Xi}}[s]^{1/2}\right\|_{2}^{2}$ and we have
\begin{small}
\begin{align}
&\left\|(\boldsymbol{\theta}^{T}[s]\otimes\boldsymbol{\theta}^{H}[s])\mathbf{U}[s]\bar{\boldsymbol{\Xi}}^{1/2}[s]\right\|_{2}^{2}=\left\|\mathrm{tr}\left(\mathbf{U}[s]\bar{\boldsymbol{\Xi}}^{1/2}[s]\tilde{\mathbf{Q}}[s]\right)\right\|_{2}^{2}.\label{proA1_3}
\end{align}
\end{small}%
It is noted that $\left\|\mathrm{tr}\left(\mathbf{U}[s]\bar{\boldsymbol{\Xi}}^{1/2}[s]\tilde{\mathbf{Q}}[s]\right)\right\|_{2}^{2}$ is convex with respect to $\mathrm{tr}\left(\mathbf{U}[s]\bar{\boldsymbol{\Xi}}^{1/2}[s]\tilde{\mathbf{Q}}[s]\right)$, so that its first-order Taylor expansion around any point $\mathrm{tr}\left(\mathbf{U}[s]\bar{\boldsymbol{\Xi}}^{1/2}[s]\tilde{\mathbf{Q}}[s]\right)$ is a lower bound of $\left\|\mathrm{tr}\left(\mathbf{U}[s]\bar{\boldsymbol{\Xi}}^{1/2}[s]\tilde{\mathbf{Q}}[s]\right)\right\|_{2}$, i.e.,
\begin{small}
\begin{align}
&\left\|\mathrm{tr}\left(\mathbf{U}[s]\bar{\boldsymbol{\Xi}}^{1/2}[s]\tilde{\mathbf{Q}}[s]\right)\right\|_{2}\geq \left\|\mathrm{tr}\left(\mathbf{U}[s]\bar{\boldsymbol{\Xi}}^{1/2}[s]\tilde{\mathbf{Q}}_{0}[s]\right)\right\|_{2}+\nonumber\\
&\frac{\partial \left\|\mathrm{tr}\left(\mathbf{U}[s]\bar{\boldsymbol{\Xi}}^{1/2}[s]\tilde{\mathbf{Q}}[s]\right)\right\|_{2}}{\partial \mathrm{tr}\left(\mathbf{U}[s]\bar{\boldsymbol{\Xi}}^{1/2}[s]\tilde{\mathbf{Q}}[s]\right)}\Bigg|_{\mathrm{tr}\left(\mathbf{U}[s]\bar{\boldsymbol{\Xi}}^{1/2}[s]\tilde{\mathbf{Q}_{0}}[s]\right)}\nonumber\\
&\left(\left\|\mathrm{tr}\left(\mathbf{U}[s]\bar{\boldsymbol{\Xi}}^{1/2}[s]\tilde{\mathbf{Q}}[s]\right)\right\|_{2}-\left\|\mathrm{tr}\left(\mathbf{U}[s]\bar{\boldsymbol{\Xi}}^{1/2}[s]\tilde{\mathbf{Q}}_{0}[s]\right)\right\|_{2}\right),\label{proA1_4}
\end{align}
\end{small}%
where
\begin{small}
\begin{align}
&\frac{\partial \left\|\mathrm{tr}\left(\mathbf{U}[s]\bar{\boldsymbol{\Xi}}^{1/2}[s]\tilde{\mathbf{Q}}[s]\right)\right\|_{2}}{\partial \mathrm{tr}\left(\mathbf{U}[s]\bar{\boldsymbol{\Xi}}^{1/2}[s]\tilde{\mathbf{Q}}[s]\right)}\Bigg|_{\mathrm{tr}\left(\mathbf{U}[s]\bar{\boldsymbol{\Xi}}^{1/2}[s]\tilde{\mathbf{Q}_{0}}[s]\right)}=\nonumber\\
&\left\|\mathrm{tr}\left(\mathbf{U}[s]\bar{\boldsymbol{\Xi}}^{1/2}[s]\tilde{\mathbf{Q}}[s]\right)\right\|_{2}^{-1}\left\|\mathrm{tr}\left(\mathbf{U}[s]\bar{\boldsymbol{\Xi}}^{1/2}[s]\tilde{\mathbf{Q}}[s]\right)\right\|_{2}.\label{proA1_5}
\end{align}
\end{small}%
Then, (\ref{proA1_5}) can be rewritten as
\begin{small}
\begin{align}
&\left\|\mathrm{tr}\left(\mathbf{U}[s]\bar{\boldsymbol{\Xi}}^{1/2}[s]\tilde{\mathbf{Q}}[s]\right)\right\|_{2}\geq\left\|\mathrm{tr}\left(\mathbf{U}[s]\bar{\boldsymbol{\Xi}}^{1/2}[s]\tilde{\mathbf{Q}}_{0}[s]\right)\right\|_{2}+\nonumber\\
&\left\|\mathrm{tr}\left(\mathbf{U}[s]\bar{\boldsymbol{\Xi}}^{1/2}[s]\tilde{\mathbf{Q}}_{0}[s]\right)\right\|_{2}^{-1}\left\|\mathrm{tr}\left(\mathbf{U}[s]\bar{\boldsymbol{\Xi}}^{1/2}[s]\tilde{\mathbf{Q}}_{0}[s]\right)\right\|_{2}\nonumber\\
&\times\left(\left\|\mathrm{tr}\left(\mathbf{U}[s]\bar{\boldsymbol{\Xi}}^{1/2}[s]\tilde{\mathbf{Q}}[s]\right)\right\|_{2}-\left\|\mathrm{tr}\left(\mathbf{U}[s]\bar{\boldsymbol{\Xi}}^{1/2}[s]\tilde{\mathbf{Q}}_{0}[s]\right)\right\|_{2}\right)\nonumber\\
&=\left\|\mathrm{tr}\left(\mathbf{U}[s]\bar{\boldsymbol{\Xi}}^{1/2}[s]\tilde{\mathbf{Q}}_{0}[s]\right)\right\|_{2}^{-1}\left\|\mathrm{tr}\left(\mathbf{U}[s]\bar{\boldsymbol{\Xi}}^{1/2}[s]\tilde{\mathbf{Q}}_{0}[s]\right)\right\|_{2}\nonumber\\
&\times\left\|\mathrm{tr}\left(\mathbf{U}[s]\bar{\boldsymbol{\Xi}}^{1/2}[s]\tilde{\mathbf{Q}}[s]\right)\right\|_{2}=\left\|\mathrm{tr}\left(\mathbf{U}[s]\bar{\boldsymbol{\Xi}}^{1/2}[s]\tilde{\mathbf{Q}}_{0}[s]\right)\right\|_{2}^{-1}\nonumber\\
&\mathrm{tr}\left(\mathbf{U}[s]\bar{\boldsymbol{\Xi}}^{1/2}[s]\tilde{\mathbf{Q}}_{0}[s]\mathbf{U}[s]\bar{\boldsymbol{\Xi}}^{1/2}[s]\tilde{\mathbf{Q}}[s]\right).\label{proA1_6}
\end{align}
\end{small}%
Then, we obtain 
\begin{small}
\begin{align}
&\mathrm{tr}(\mathbf{C}_{1}[s]\tilde{\mathbf{Q}}[s]\mathbf{C}_{2}[s]\tilde{\mathbf{Q}}[s])=\left\|\mathrm{tr}\left(\mathbf{U}[s]\bar{\boldsymbol{\Xi}}^{1/2}[s]\tilde{\mathbf{Q}}[s]\right)\right\|_{2}^{2}\geq\nonumber\\
& \left\|\mathrm{tr}\left(\mathbf{U}[s]\bar{\boldsymbol{\Xi}}^{1/2}[s]\tilde{\mathbf{Q}}_{0}[s]\right)\right\|_{2}^{-2}\mathrm{tr}\left(\mathbf{U}[s]\bar{\boldsymbol{\Xi}}^{1/2}[s]\tilde{\mathbf{Q}}_{0}[s]\mathbf{U}[s]\bar{\boldsymbol{\Xi}}^{1/2}[s]\tilde{\mathbf{Q}}[s]\right)^{2}.\label{proA1_7}
\end{align}
\end{small}%
which completes the proof.

\section{Proof of Rank One}\label{appB}
According to \cite{boyd2004convex}, the epigraph of problem (\ref{pro54}) is expressed as
\begin{small}
\begin{subequations}
\begin{align}
\min_{\tilde{\mathbf{Q}}[s]\succeq \mathbf{0}}&~-\log_{2}\left(\mathrm{tr}((\mathbf{F}_{U}[s]+\mathbf{F}_{E}[s])\tilde{\mathbf{Q}}[s])+\sigma_{u}^{2}\right)-\nonumber\\
&\log_{2}\left(\mathrm{tr}(\mathbf{F}_{T}[s]\tilde{\mathbf{Q}}[s])+\sigma_{u}^{2}\right)
-\left(\log_{2}\left(\mathrm{tr}(\mathbf{F}_{E}[s]\tilde{\mathbf{Q}}_{0}[s])+\sigma_{u}^{2}\right)\right.\nonumber\\
&+\log_{2}\left(\mathrm{tr}((\mathbf{F}_{U}[s]+\mathbf{F}_{T}[s])\right.\left.\left.\tilde{\mathbf{Q}}_{0}[s])+\sigma_{u}^{2}\right)\right)\nonumber\\
&-\frac{1}{\ln 2(\mathrm{tr}(\mathbf{F}_{E}[s]\tilde{\mathbf{Q}}_{0}[s])+\sigma_{u}^{2})}\mathrm{tr}(\mathbf{F}_{E}[s](\tilde{\mathbf{Q}}[s]-\tilde{\mathbf{Q}}_{0}[s]))\nonumber\\
&-\frac{1}{\ln 2(\mathrm{tr}((\mathbf{F}_{T}[s]+\mathbf{F}_{U}[s])\tilde{\mathbf{Q}}_{0}[s])+\sigma_{u}^{2})}\times\nonumber\\
&\mathrm{tr}((\mathbf{F}_{T}[s]+\mathbf{F}_{U}[s])(\tilde{\mathbf{Q}}[s]-\tilde{\mathbf{Q}}_{0}[s])),\label{proA2_1}\\
\mbox{s.t.}~
&\mathrm{tr}((\mathbf{F}_{U}[s]+\mathbf{F}_{E}[s])\tilde{\mathbf{Q}}[s])\geq\eta_{1},&\label{proA2_2}\\
&\mathrm{tr}(\mathbf{F}_{T}[s]\tilde{\mathbf{Q}}[s])\geq\eta_{2},&\label{proA2_3}\\
&\mathrm{tr}\left(\mathbf{B}_{i}\tilde{\mathbf{Q}}[s]\right)=1,&\label{proA2_4}\\
&\left\|\mathrm{tr}\left(\mathbf{U}[s]\bar{\boldsymbol{\Xi}}^{1/2}[s]\tilde{\mathbf{Q}}_{0}[s]\right)\right\|_{2}^{-1}\nonumber\\
&\times\mathrm{tr}\left(\mathbf{U}\bar{\boldsymbol{\Xi}}^{1/2}\tilde{\mathbf{Q}}_{0}[s]\mathbf{U}[s]\bar{\boldsymbol{\Xi}}^{1/2}[s]\tilde{\mathbf{Q}}[s]\right)\geq \sqrt{\gamma_{s}\sigma_{b}^{2}},&\label{proA2_5}\\
&\tilde{\mathbf{Q}}[s]\succeq\mathbf{0}.&\label{proA2_6}
\end{align}\label{proA2_7}%
\end{subequations}
\end{small}%
Since $\tilde{\mathbf{Q}}_{k}[s]$ optimization problem is convex, the Slater's condition is satisfied \cite{boyd2004convex}. 
Consequently, the duality gap between the primal and dual problems is zero. By solving the dual problem, we can obtain the optimal solution for this problem. Let $\mathbf{Q}_{k}[s]=\boldsymbol{\theta}_{k}[s]\boldsymbol{\theta}_{k}^{H}[s],~\forall~k$, the Lagrangian function corresponding to this problem can be defined as 
\begin{small}
\begin{align}
&\mathcal{L}=-\log_{2}\left(\eta_{1}+\sigma_{u}^{2}\right)-\log_{2}\left(\eta_{2}+\sigma_{u}^{2}\right)\nonumber\\
&-\left(\log_{2}\left(\mathrm{tr}(\mathbf{F}_{E}[s]\tilde{\mathbf{Q}}_{0}[s])+\sigma_{u}^{2}\right)+\right.\nonumber\\
&\left.\log_{2}\left(\mathrm{tr}((\mathbf{F}_{U}[s]+\mathbf{F}_{T}[s])\tilde{\mathbf{Q}}_{0}[s])+\sigma_{u}^{2}\right)\right)\nonumber\\
&-\frac{1}{\ln 2(\mathrm{tr}(\mathbf{F}_{E}[s]\tilde{\mathbf{Q}}_{0}[s])+\sigma_{u}^{2})}\mathrm{tr}(\mathbf{F}_{E}[s](\tilde{\mathbf{Q}}[s]-\tilde{\mathbf{Q}}_{0}[s]))\nonumber\\
&-\frac{1}{\ln 2(\mathrm{tr}((\mathbf{F}_{T}[s]+\mathbf{F}_{U}[s])\tilde{\mathbf{Q}}_{0}[s])+\sigma_{u}^{2})}\mathrm{tr}((\mathbf{F}_{T}[s]+\mathbf{F}_{U}[s])(\tilde{\mathbf{Q}}[s]\nonumber\\
&-\tilde{\mathbf{Q}}_{0}[s]))+\lambda_{1}\left(\left\|\mathrm{tr}\left(\mathbf{U}[s]\bar{\boldsymbol{\Xi}}^{1/2}[s]\tilde{\mathbf{Q}}_{0}[s]\right)\right\|_{2}^{-1}\right.\nonumber\\
&\left.\mathrm{tr}\left(\mathbf{U}[s]\bar{\boldsymbol{\Xi}}^{1/2}[s]\tilde{\mathbf{Q}}_{0}[s]\mathbf{U}[s]\bar{\boldsymbol{\Xi}}^{1/2}[s]\tilde{\mathbf{Q}}[s]\right)-\sqrt{\gamma_{s}\sigma_{b}^{2}}\right)-\mathrm{tr}\left(\boldsymbol{\Gamma}\tilde{\mathbf{Q}}[s]\right)\nonumber\\
&+\lambda_{2,i}\sum_{i=1}^{N}\left(\mathrm{tr}\left(\mathbf{B}_{i}\tilde{\mathbf{Q}}[s]\right)-1\right)+\lambda_{3}\left(\mathrm{tr}((\mathbf{F}_{U}[s]+\mathbf{F}_{E}[s])\tilde{\mathbf{Q}}[s])\right.\nonumber\\-&\left.\eta_{1}\right)+\lambda_{4}\left(\mathrm{tr}(\mathbf{F}_{T}[s]\tilde{\mathbf{Q}}[s])-\eta_{2}\right).&\label{proA2_8}
\end{align}
\end{small}%
To simplify this formula (\ref{proA2_2}) and $\mathcal{L}$ is rewritten as
\begin{small}
\begin{align}
&\mathcal{L}=-\frac{1}{\ln 2(\mathrm{tr}(\mathbf{F}_{E}[s]\tilde{\mathbf{Q}}_{0}[s])+\sigma_{u}^{2})}\mathrm{tr}(\mathbf{F}_{E}[s]\tilde{\mathbf{Q}}[s])\nonumber\\
&-\frac{1}{\ln 2(\mathrm{tr}((\mathbf{F}_{T}[s]+\mathbf{F}_{U}[s])\tilde{\mathbf{Q}}_{0}[s])+\sigma_{u}^{2})}\mathrm{tr}((\mathbf{F}_{T}[s]+\mathbf{F}_{U}[s])\tilde{\mathbf{Q}}[s])\nonumber\\
&+\lambda_{1}\left(\left\|\mathrm{tr}\left(\mathbf{U}[s]\bar{\boldsymbol{\Xi}}^{1/2}[s]\tilde{\mathbf{Q}}_{0}[s]\right)\right\|_{2}^{-1}\right.\nonumber\\
&\left.\mathrm{tr}\left(\mathbf{U}[s]\bar{\boldsymbol{\Xi}}^{1/2}[s]\tilde{\mathbf{Q}}_{0}[s]\mathbf{U}[s]\bar{\boldsymbol{\Xi}}^{1/2}[s]\tilde{\mathbf{Q}}[s]\right)\right)-\mathrm{tr}\left(\boldsymbol{\Gamma}\tilde{\mathbf{Q}}[s]\right)\nonumber\\
&+\lambda_{2,i}\sum_{i=1}^{N}\mathrm{tr}\left(\mathbf{B}_{i}\tilde{\mathbf{Q}}[s]\right)+\lambda_{3}\mathrm{tr}((\mathbf{F}_{U}[s]+\mathbf{F}_{E}[s])\tilde{\mathbf{Q}}[s])\nonumber\\
&+\lambda_{4}\mathrm{tr}(\mathbf{F}_{T}[s]\tilde{\mathbf{Q}}[s])+\gamma,&\label{proA2_9}
\end{align}
\end{small}%
where $\lambda_{1}\geq 0$, $\lambda_{2}\geq 0$, $\lambda_{3}\geq 0$ and $\lambda_{4}\geq 0$ is Lagrangian factor. $\boldsymbol{\Gamma}\succeq\mathbf{0}$ is Lagrangian multiplier matrix. $\gamma$ includes all terms that do not involve $\tilde{\mathbf{Q}}[s]$, and $\gamma$ is expressed as
\begin{small}
\begin{align}
&\gamma=-\log_{2}\left(\eta_{1}+\sigma_{u}^{2}\right)-\log_{2}\left(\eta_{2}+\sigma_{u}^{2}\right)+\nonumber\\
&\log_{2}\left(\mathrm{tr}((\mathbf{F}_{U}[s]+\mathbf{F}_{T}[s])\tilde{\mathbf{Q}}_{0}[s])+\sigma_{u}^{2}\right)+\frac{1}{\ln 2(\mathrm{tr}(\mathbf{F}_{E}[s]\tilde{\mathbf{Q}}_{0}[s])+\sigma_{u}^{2})}\nonumber\\
&\mathrm{tr}(\mathbf{F}_{E}[s]\tilde{\mathbf{Q}}_{0}[s])+\frac{1}{\ln 2(\mathrm{tr}((\mathbf{F}_{T}[s]+\mathbf{F}_{U}[s])\tilde{\mathbf{Q}}_{0}[s])+\sigma_{u}^{2})}\nonumber\\
&\times\mathrm{tr}((\mathbf{F}_{T}[s]+\mathbf{F}_{U}[s])\tilde{\mathbf{Q}}_{0}[s])+\lambda_{1}K-\lambda_{2}\sqrt{\gamma_{s}\sigma_{b}^{2}}-\lambda_{4}\eta_{1}-\lambda_{5}\eta_{2}.\label{proA2_10}
\end{align}
\end{small}%
According to the Karush-Kuhn-Tucker(KKT) conditions theory, we have
\begin{small}
\begin{align}
&\boldsymbol{\Gamma}[s]^{*}\tilde{\mathbf{Q}}[s]^{*}=\mathbf{0}, \lambda_{1}^{*}\geq 0, \lambda_{2,i}^{*}\geq 0, \lambda_{3}^{*}\geq 0, \lambda_{4}^{*}\geq 0,\frac{\partial \mathcal{L}}{\partial \tilde{\mathbf{Q}}[s]}\Bigg|_{\tilde{\mathbf{Q}}[s]^{*}}=\mathbf{0},\label{proA2_11}
\end{align}
\end{small}%
where $\frac{\partial \mathcal{L}}{\partial \tilde{\mathbf{Q}}[s]}\Bigg|_{\tilde{\mathbf{Q}}[s]^{*}}$ is expressed as
\begin{small}
\begin{align}
&-\underbrace{\frac{1}{\ln 2(\mathrm{tr}(\mathbf{F}_{E}[s]\tilde{\mathbf{Q}}_{0}[s])+\sigma_{u}^{2})}}_{\rho_{1}}\mathbf{F}_{E}[s]\nonumber\\
&-\underbrace{\frac{1}{\ln 2(\mathrm{tr}((\mathbf{F}_{T}[s]+\mathbf{F}_{U}[s])\tilde{\mathbf{Q}}_{0}[s])+\sigma_{u}^{2})}}_{\rho_{2}}(\mathbf{F}_{T}[s]+\mathbf{F}_{U}[s])\nonumber
\end{align}
\end{small}%
\begin{small}
\begin{align}
&+\lambda_{1}\underbrace{\left(\left\|\mathrm{tr}\left(\mathbf{U}[s]\bar{\boldsymbol{\Xi}}^{1/2}[s]\tilde{\mathbf{Q}}_{0}[s]\right)\right\|_{2}^{-1}\mathbf{U}[s]\bar{\boldsymbol{\Xi}}^{1/2}[s]\tilde{\mathbf{Q}}_{0}[s]\mathbf{U}[s]\bar{\boldsymbol{\Xi}}^{1/2}[s]\right)}_{\mathbf{M}[s]}\nonumber\\
&-\boldsymbol{\Gamma}+\mathrm{diag}(\boldsymbol{\lambda}_{2})\mathbf{I}+\lambda_{3}(\mathbf{F}_{U}[s]+\mathbf{F}_{E}[s])+\lambda_{4}\mathbf{F}_{T}[s]+\gamma=\mathbf{0},&\label{proA2_12}
\end{align}
\end{small}%
where $\boldsymbol{\lambda}=\left[\lambda_{2,1},\ldots,\lambda_{2,N}\right]$. Based on (\ref{proA2_12}), we have (\ref{proA2_13}) at the top of next page.
\begin{figure*}[ht]
\begin{small}
\begin{align}
&\boldsymbol{\Gamma}^{*}[s]=\mathrm{diag}(\boldsymbol{\lambda}_{2})\mathbf{I}-\underbrace{\left(\rho_{1}\mathbf{F}_{E}[s]+\rho_{2}(\mathbf{F}_{T}[s]+\mathbf{F}_{U}[s])-\lambda_{1}\mathbf{M}-\lambda_{3}(\mathbf{F}_{U}[s]+\mathbf{F}_{E}[s])-\lambda_{4}\mathbf{F}_{T}[s]\right)}_{\boldsymbol{\Delta}[s]}\nonumber\\
&=\mathrm{diag}(\boldsymbol{\lambda}_{2})\mathbf{I}-\boldsymbol{\Delta}[s].\label{proA2_13}
\end{align}
\end{small}%
\hrulefill
\end{figure*}
Next, $\tilde{\mathbf{Q}}^{*}[s]$ is rank-one and will be proved by analyzing its structure. Firstly, we define $\xi_{max}$ as the maximum eigenvalue of $\boldsymbol{\Gamma}[s]^{*}$. Notably, due to the random nature of the channel, the probability of multiple eigenvalues having the same maximum value is zero. In accordance with \cite{9133130}, if $\xi_{max}>1$, $\boldsymbol{\Gamma}^{*}[s]$ is not positive semidefinite, it contradicts $K_{1}$. Consequently, $\xi_{max}<1$, $\boldsymbol{\Gamma}^{*}[s]$ must be positive definite and full rank. Based on  $K_{2}$, it is apparent that $\tilde{\mathbf{Q}}^{*}[s]$ can only be $\mathbf{0}$, which contradicts reality. Thus, $\xi_{max}<1$ must be satisfied, which implies $\mathrm{Rank}(\boldsymbol{\Gamma}^{*}[s])=N-1$. Thus, $\mathrm{Rank}(\tilde{\mathbf{Q}}^{*}[s])=1$, i.e., $\tilde{\mathbf{Q}}^{*}[s]$ is determined to be of rank-one. This completes the proof.
\end{document}